\documentclass[letterpaper,twocolumn,10pt]{article}

\usepackage[numbers,sort,compress]{natbib}

\usepackage{hyperref}
\usepackage{breakurl}
\usepackage{oneinchmargins}
\usepackage{tightenum}
\usepackage{wrapfig}
\usepackage{pifont}

\usepackage{enumitem}
\usepackage{listings}
\usepackage{latexsym}
\usepackage{adjustbox}
\usepackage{graphicx}
\usepackage{floatflt}
\usepackage{setspace}
\usepackage{graphicx}
\usepackage{algorithm}
\usepackage{amsmath}
\usepackage[small,compact]{titlesec}
\usepackage{titling}

\usepackage{xspace}
\usepackage{makecell}
\usepackage{caption}

\usepackage[normalem]{ulem}

\newcommand{\ke}[1]{{\color{pink}Ke:  {#1}}}
\newcommand{\theo}[1]{{\color{blue}Theo: \textbf{#1}}}
\newcommand{\theonote}[1]{}

\newcommand{\yiying}[1]{{\color{red}Yiying: \textbf{#1}}}

\newcommand{\nonapproximate}[1]{non-approximate\xspace}

\renewcommand{\em}{\it}

\newcommand{\comment}[1]{}
\newcommand{\ignore}[1]{}

\newcommand{\boldparagraph}[1]{\vspace*{1ex}\noindent\textbf{#1}\hspace{1em}}

\def\cfigure[#1,#2,#3]{
\begin{figure}
\vspace*{0mm}
\begin{center}

\includegraphics[width=3in]{#1} 
 
\vspace*{-3mm}\caption[]{#2
} \label{#3}
 
\vspace*{-5mm}
\end{center}
\end{figure}}

\def\cfigurefour[#1,#2,#3]{
\begin{figure}
\vspace*{0mm}
\begin{center}

\includegraphics[width=4in]{#1} 
 
\vspace*{-3mm}\caption[]{#2
} \label{#3}
 
\vspace*{-5mm}
\end{center}
\end{figure}}

\def\cfiguretemp[#1,#2,#3]{
\begin{figure}
\vspace*{0mm}
\begin{center}

\includegraphics[width=3.5in]{#1} 
 
\vspace*{-3mm}\caption[]{#2
} \label{#3}
 
\vspace*{-5mm}
\end{center}
\vspace*{-2mm}
\end{figure}}

\def\wfigure[#1,#2,#3]{
\begin{figure*}
\vspace*{0mm}
\begin{center}
 \includegraphics[width=\textwidth]{#1} 
 \vspace*{-3mm}\caption[]{#2
} \label{#3}
 
\end{center}
\end{figure*}}

\def\threefigure[#1,#2,#3,#4,#5]{
\begin{figure*}
\vspace*{0mm}
\begin{center}

\begin{tabular}{ccc}
\includegraphics[width=2in]{#1} & \includegraphics[width=2in]{#2} &  \includegraphics[width=2in]{#3} \\
(a) & (b) & (c) \\
\end{tabular}

\vspace*{-3mm}\caption[]{#4
} \label{#5}

\vspace*{-5mm}
\end{center}
\vspace*{-2mm}
\end{figure*}}

\def\dcfigure[#1,#2,#3,#4,#5,#6]{
{
\begin{figure*}
\begin{center}
\begin{minipage}[c]{\columnwidth}{
\includegraphics[width=\columnwidth]{#1} 
\vspace*{0mm}\caption[]{#2} \label{#3} \
}\end{minipage}\hspace*{\columnsep}\
\begin{minipage}[c]{\columnwidth}{
\includegraphics[width=\columnwidth]{#4} 
\vspace*{0mm}\caption[]{#5}\label{#6} \
}\end{minipage}
\end{center}
\end{figure*}
}
}

\def\tableByTable[#1,#2,#3,#4,#5,#6]{
{
\begin{table*}
\begin{center}
\begin{minipage}[c]{3in}{
\centering
{#1}
\vspace*{0mm}\tabcaption[]{#2}\label{#3} \
}\end{minipage}\hspace*{\columnsep}\
\begin{minipage}[c]{3in}{
\centering
{#4}
\vspace*{0mm}\tabcaption[]{#5}\label{#6} \
}\end{minipage}
\end{center}
\end{table*}
}
}

\def\figureByTable[#1,#2,#3,#4,#5,#6]{
{
\begin{figure*}
\begin{center}
\begin{minipage}[c]{3in}{
\centering
\includegraphics[width=\textwidth]{#1}
\vspace*{0mm}\figcaption[]{#2} \label{#3} \
}\end{minipage}\hspace*{\columnsep}\
\begin{minipage}[c]{3.3in}{
\centering
{#4}
\vspace*{0mm}\tabcaption[]{#5}\label{#6} \
}\end{minipage}
\end{center}
\end{figure*}
}
}

\def\tableByFigure[#1,#2,#3,#4,#5,#6]{
{
\begin{figure*}
\begin{center}
\begin{minipage}[c]{4.3in}{
\centering
{#1}
\vspace*{0mm}\tabcaption[]{#2} \label{#3} \
}\end{minipage}\hspace*{\columnsep}\
\begin{minipage}[c]{2.2in}{
\centering
\includegraphics[width=\textwidth]{#4}
\vspace*{-0.35in}\caption[]{#5}\label{#6} \
}\end{minipage}
\end{center}
\end{figure*}
}
}

\def\doublecfigure[#1,#2,#3,#4]{
{
\begin{figure}
\begin{center}
\begin{minipage}[c]{1.5in}{
\begin{center}
\includegraphics[width=1.5in]{#1}
\end{center}
}\end{minipage}\hspace*{1em}\
\begin{minipage}[c]{1.5in}{
\begin{center}
\includegraphics[width=1.5in]{#2}
\end{center}
}\end{minipage}
\vspace*{0mm}\caption[]{#3} \label{#4} \
\end{center}
\end{figure}
}
}

\def\qcfigure[#1,#2,#3,#4,#5,#6]{
{
\begin{figure*}
\vspace*{0.2in}\
\begin{center}
\begin{minipage}[c]{3in}{
\includegraphics[width=3in]{#1} 
\vspace*{-3mm}
}
\end{minipage}\hspace*{0.5in}\
\begin{minipage}[c]{3in}{
\includegraphics[width=3in]{#2} 
\vspace*{-3mm}
}\end{minipage}

\begin{minipage}[c]{3in}{
\includegraphics[width=3in]{#3} 
\vspace*{-3mm}
}
\end{minipage}\hspace*{0.5in}\
\begin{minipage}[c]{3in}{
\includegraphics[width=3in]{#4} 
\vspace*{-3mm}
}\end{minipage}
\end{center}
\caption[]{#5}\label{#6}
\end{figure*}
}
}

\def\twfigure[#1,#2,#3,#4,#5]{
{
\begin{figure*}
\vspace*{0.2in}\
\begin{center}
\begin{minipage}[c]{6.5in}{
\includegraphics[width=6.5in]{#1} 
\vspace*{-3mm}
}
\end{minipage}

\begin{minipage}[c]{6.5in}{
\includegraphics[width=6.5in]{#2} 
\vspace*{-3mm}
}\end{minipage}

\begin{minipage}[c]{6.5in}{
\includegraphics[width=6.5in]{#3} 
\vspace*{-3mm}
}
\end{minipage}
\end{center}
\caption[]{#4}\label{#5}
\end{figure*}
}
}

\def\dwfigure[#1,#2,#3,#4]{
{
\begin{figure*}
\vspace*{0.2in}\
\begin{center}
\begin{minipage}[c]{6.5in}{
\includegraphics[width=6.5in]{#1} 
\vspace*{-3mm}
}
\end{minipage}

\begin{minipage}[c]{6.5in}{
\includegraphics[width=6.5in]{#2} 
\vspace*{-3mm}
}\end{minipage}

\end{center}
\caption[]{#3}\label{#4}
\end{figure*}
}
}

\def\dssfigure[#1,#2,#3,#4,#5,#6]{
{
\begin{figure*}
\vspace*{0.2in}\
\begin{center}
\begin{minipage}[c]{4in}{
\includegraphics[width=4in]{#1}
\vspace*{-3mm}\caption[]{#2} \label{#3} \
}\end{minipage}\hspace*{0.5in}\
\begin{minipage}[c]{2in}{
\includegraphics[width=2in]{#4}
\vspace*{-3mm}\caption[]{#5}\label{#6} \
}\end{minipage}
\end{center}
\vspace*{-0.4in}\
\end{figure*}
}
}

\def\dsfigure[#1,#2,#3,#4,#5,#6]{
{
\begin{figure*}
\vspace*{0.2in}\
\begin{center}
\begin{minipage}[c]{3in}{
\includegraphics[width=3in]{#1}
\vspace*{-3mm}\caption[]{#2} \label{#3} \
}\end{minipage}\hspace*{0.5in}\
\begin{minipage}[c]{3in}{
\hspace*{0.5in}\
\includegraphics[height=3in]{#4}
\vspace*{-3mm}\caption[]{#5}\label{#6} \
}\end{minipage}
\end{center}
\vspace*{-0.4in}\
\end{figure*}
}
}

\def\dsyfigure[#1,#2,#3,#4,#5,#6]{
{
\begin{figure*}
\vspace*{0.2in}\
\begin{center}
\begin{minipage}[c]{2.5in}{
\includegraphics[height=2.5in]{#1}
\vspace*{-3mm}\caption[]{#2} \label{#3} \
}\end{minipage}\hspace*{0.5in}\
\begin{minipage}[c]{2.5in}{
\includegraphics[height=2.5in]{#4}
\vspace*{-3mm}\caption[]{#5}\label{#6} \
}\end{minipage}
\end{center}
\vspace*{-0.4in}\
\end{figure*}
}
}

\def\dyfigure[#1,#2,#3,#4,#5,#6]{
{
\begin{figure*}
\vspace*{0.2in}\
\begin{center}
\begin{minipage}[c]{3in}{
\includegraphics[height=3in]{#1} 
\vspace*{-3mm}\caption[]{#2} \label{#3} \
}\end{minipage}\hspace*{0.5in}\
\begin{minipage}[c]{3in}{
\includegraphics[height=3in]{#4} 
\vspace*{-3mm}\caption[]{#5}\label{#6} \
}\end{minipage}
\end{center}
\vspace*{-0.4in}\
\end{figure*}
}
}

\def\dyoldfigure[#1,#2,#3,#4,#5,#6]{
{
\begin{figure*}
\vspace*{0.2in}\
\begin{center}
\begin{minipage}[c]{3in}{
\epsfysize=2.0in\
\hspace{0.5in}\
\epsfbox{#1}
\vspace*{-3mm}\caption[]{#2} \label{#3} \
}\end{minipage}\hspace*{0.25in}\
\begin{minipage}[c]{3in}{
\epsfysize=2.0in\
\hspace{0.5in}\
\epsfbox{#4}
\vspace*{-3mm}\caption[]{#5}\label{#6} \
}\end{minipage}
\end{center}
\vspace*{-0.4in}\
\end{figure*}
}
}

\def\cfiguredouble[#1,#2,#3,#4]{
\begin{figure}
\vspace*{0.2in}\
\begin{center}
\begin{minipage}[c]{1.5in}{
\epsfxsize=1.5in\
\epsfbox{#1}
}\end{minipage}\hspace*{0.1in}\
\begin{minipage}[c]{1.5in}{
\epsfxsize=1.5in\
\vspace{0.1in}\epsfbox{#2}
}\end{minipage}\vspace*{-0.10in} \caption[]{#3}\label{#4}
\end{center}
\vspace*{-0.4in}\
\end{figure}
}

\def\wpfigure[#1,#2,#3,#4]{
\begin{figure*}
\vspace*{4mm}
\begin{center}

\includegraphics[width=#4]{#1} 

\vspace*{-3mm}\caption[]{#2
} \label{#3}

\vspace*{-5mm}
\end{center}
\end{figure*}}

\def\wprfigure[#1,#2,#3,#4,#5]{
\begin{figure*}
\vspace*{4mm}
\begin{center}

\includegraphics[width=#4, angle=#5]{#1} 

\vspace*{-3mm}\caption[]{#2
} \label{#3}

\vspace*{-5mm}
\end{center}
\end{figure*}}

\def\DoubleFigureWSlide[#1,#2,#3,#4,#5,#6,#7,#8,#9]{
\begin{figure*}
\vspace*{#9}
\begin{center}
\begin{minipage}{#4}
\includegraphics[width=#4]{#1}
\vspace*{-3mm}\caption{#2
}\label{#3}
\end{minipage}
\hspace{2em}
\begin{minipage}{#8}
\includegraphics[width=#8]{#5}
\vspace*{-3mm}\caption{#6
}\label{#7}
\end{minipage}
\vspace*{-5mm}
\end{center}
\end{figure*}
}

\def\DoubleFigureW[#1,#2,#3,#4,#5,#6,#7,#8]{
\begin{figure*}
\vspace*{0in}
\begin{center}
\begin{minipage}{#4}
\includegraphics[width=#4]{#1}
\vspace*{-3mm}\caption{#2
}\label{#3}
\end{minipage}
\hspace{2em}
\begin{minipage}{#8}
\includegraphics[width=#8]{#5}
\vspace*{-3mm}\caption{#6
}\label{#7}
\end{minipage}
\vspace*{-5mm}
\end{center}
\end{figure*}
}

\def\DoubleFigureWHack[#1,#2,#3,#4,#5,#6,#7,#8]{
\begin{figure*}
\vspace*{0in}
\begin{center}
\begin{minipage}{3in}
\includegraphics[width=#4]{#1}
\vspace*{-3mm}\caption{#2
}\label{#3}
\end{minipage}
\hspace{2em}
\begin{minipage}{3in}
\includegraphics[width=#8]{#5}
\vspace*{-3mm}\caption{#6
}\label{#7}
\end{minipage}
\vspace*{-5mm}
\end{center}
\end{figure*}
}

\def\ddcfigure[#1,#2,#3,#4]{
\begin{figure*}
\vspace*{0.2in}\
\begin{center}
\begin{minipage}[c]{\columnwidth}{
\includegraphics[width=\columnwidth]{#1} 
}\end{minipage}\hspace{0.5in}\
\begin{minipage}[c]{\columnwidth}{
\includegraphics[width=\columnwidth]{#2} 
}\end{minipage} \caption[]{#3}\label{#4}
\end{center}
\end{figure*}
}

\def\ddcfigureSlide[#1,#2,#3,#4,#5]{
\begin{figure*}
\vspace*{#5}\
\begin{center}
\begin{minipage}[c]{3in}{
\includegraphics[height=3in]{#1} 
}\end{minipage}\hspace{0.5in}\
\begin{minipage}[c]{3in}{
\includegraphics[height=3in]{#2} 
}\end{minipage}\vspace*{-0.10in} \caption[]{#3}\label{#4}
\end{center}
\vspace*{-0.4in}\
\end{figure*}
}

\def\cxfigure[#1,#2,#3]{
\begin{figure}
\vspace*{4mm}
\begin{center}
 
\epsfxsize=2.5in\
\epsfbox{#1}\
 
\vspace*{-0.10in}\caption[]{#2
} \label{#3}
 
\vspace*{-5mm}
\end{center}
\vspace*{-2mm}
\end{figure}}

\newcommand{\beforecaption}{\vspace{-.15cm}\begin{spacing}{0.85}}
\newcommand{\aftercaption}{\vspace{-.45cm}\end{spacing}}
\newcommand{\mycaption}[3]{\beforecaption\caption{\label{#1}{\bf \footnotesize #2} \em\small #3}\aftercaption}

\newcommand{\eg}{\textit{e.g.}}
\newcommand{\ie}{\textit{i.e.}}

\newcommand{\MB}{\,MB}
\newcommand{\GB}{\,GB}

\newcommand{\gbps}{\,Gbps}

\newcommand{\atpName} {ATP\xspace}
\newcommand{\atp}{\atpName}

\newcommand{\atpbase}{ATP\_Base}
\newcommand{\atprc}{ATP\_RC}

\newcommand{\atpfull}{ATP\_Full}
\newcommand{\atpsched}{ATP\_MRDF}
\newcommand{\atpfixtlr}{ATP\_FixTLR}
\newcommand{\atpfixqueue}{ATP\_FixPrio}

\newcommand{\mlr}{\texttt{MaxLossRate}\xspace}
\newcommand{\msr}{$MSR$\xspace}
\newcommand{\tlr}{$TLR$\xspace}

\newcommand{\earlysend}{EarlySend}
\newcommand{\midsend}{SampleSend}

\newcommand{\sysname}{NetApprox}

\newcommand{\jct}{JCT}

\newcommand{\msgid}{\texttt{MsgID}\xspace}

\newcommand{\nextmsgid}{\texttt{NextMsgID}\xspace}

\newcommand{\seq}{\texttt{SeqNum}}

\newcommand{\dctcp}{DCTCP\xspace}

\newcommand{\preload}{\texttt{LD\_PRELOAD}}
\newcommand{\approxlib}{\mbox{ApproxLib}\xspace}
\newcommand{\tauhat}{\widehat\tau}
\newcommand{\Writev}{\texttt{writev}}
\newcommand{\Write}{\texttt{write}}

\newcommand{\FBKV}[1]{FBKV\xspace}
\newcommand{\FBHD}[1]{FBHD\xspace}
\newcommand{\avg}{\texttt{Avg}}
\newcommand{\rank}{\texttt{Rank}}

\usepackage{xcolor}
\definecolor{commentgreen}{RGB}{2,112,10}
\definecolor{eminence}{RGB}{108,48,130}
\definecolor{weborange}{RGB}{255,165,0}
\definecolor{frenchplum}{RGB}{129,20,83}

\usepackage{listings}
\usepackage{geometry}
\begin{document}

\pagestyle{plain}

\title{\Large\bf Exploiting Network Loss for Distributed Approximate Computing with \sysname}

\author{{\rm Ke Liu\textsuperscript{\#}\thanks{Most work done while at Purdue University}} \qquad 
{\rm Jinmou Li\textsuperscript{\textparagraph}} \qquad 
{\rm Shin-Yeh Tsai\textsuperscript{$\mathsection$}\thanks{Work done while at Purdue University}} \qquad 
{\rm Theophilus Benson\textsuperscript{\textdaggerdbl}} \qquad 
{\rm Yiying Zhang\textsuperscript{\textparagraph}}
\\ \\ {\textsuperscript{\#}Institute of Computing Technology} \qquad 
{\textsuperscript{\textparagraph}UC San Diego} \qquad 
{\textsuperscript{$\mathsection$}Facebook} \qquad
{\textsuperscript{\textdaggerdbl}Brown University} \qquad
}

\maketitle

\thispagestyle{plain}

\section*{Abstract}
Many data center applications such as machine learning and big data analytics can complete their analysis without processing the complete set of data.
While extensive approximate-aware optimizations have been proposed
at hardware, programming language, and application levels; however, to date, the approximate computing optimizations have ignored the network layer. 
%

We propose \sysname, which to the best of our knowledge, is the first approximate-aware network layer comprising transport-layer protocol, network resource allocation schemes, and scheduling/priority-assignment policies.   
Building on the observation that approximate applications can tolerate loss, \sysname's main insights are to aggressively send approximate traffic (which improves 
the performance of approximate applications)
and to minimize the network resources allocated to approximate traffic (which simultaneously limits the impact of aggressive approximate traffic while freeing up resources that, in turn, improve non-approximate applications' performance).
We ported Flink, Kafka, Spark, and PyTorch to \sysname\ and evaluated \sysname\ with both large-scale simulation and real implementation. 
Our evaluation results show that \sysname\ improves job completion times by up to 80\%\ compared to network-oblivious approximation solutions, and improves the performance of co-running non-approximate workloads by 79\%.

\section{Introduction}
Today, big-data workloads, such as stream data processing, data analytics, and machine learning training, dominate network communication in data centers and clouds. Despite significant efforts in designing and tailoring data-center networks to optimize their performance
~\cite{PIAS-nsdi15, Grosvenor-nsdi2015, Munir-Sigcomm14, Changhyun-ATC2015, Cho17-SIGCOMM, ndp-sigcomm17, Homa-Sigcomm18, Hermes-sigcomm17, letitflow-nsdi17, ACC-sigcomm21, Powertcp-nsdi22, Aeolus-sigcomm20, Swift-sigcomm20, Backpressure-nsdi22, Hpcc-sigcomm19, Wei-nsdi16, Bai-CoNext16}
network systems continue to overlook a key property of these big-data workloads: their approximate nature. These workloads (often called \emph{approximate computing} workloads) are capable of completing a job with incomplete or imprecise input data or intermediate data, a property we call {\em slack tolerance}. For example, ML training can achieve high accuracy even without certain (\eg, tiny or zero) gradients when updating weights~\cite{lin2018deep, Casync-sosp21}. Similarly, data-analytics jobs like MapReduce workloads can achieve satisfactory accuracy even when some input data or intermediate data between mappers and reducers are lost~\cite{Goiri-Asplos15, StreamApprox2017}. 

Recent efforts to leverage approximation slack tolerance have demonstrated significant performance and energy benefits. 
However, these efforts have focused on the programming languages~\cite{Baek10-PLDI, Bornholt-ASPLOS14,Sampson11-PLDI, Sampson13-MICRO}, 
hardware~\cite{Sampson13-MICRO}, and application~\cite{Agarwal13-EUROSYS, Ananthanarayanan14-NSDI, Chaudhuri07-IEEE,Condie10-NSDI, Goiri-Asplos15, Hellerstein97-SIGMOD, Pansare11-PVLDB,  StreamApprox2017,Chris-VLDB08} levels, all within a single-server setting.
However, most data-center approximate applications run in a distributed manner,
and it is imperative to exploit slack tolerance in a distributed setting. 
Existing distributed approximate applications have been treating the network as a blackbox by either dropping some data before sending the rest to the network~\cite{Goiri-Asplos15} or sending all the data and then dropping some at the receiver before the computation~\cite{StreamApprox2017}.
These existing solutions reuse existing networking protocols, scheduling principles, and queuing disciplines, missing the opportunities to potentially incorporate slack tolerance {\em at the network layer}. 

Motivated by this observation, our work aims to answer these questions: 
``Instead of treating the network as a black box, 
\textbf{is it beneficial to make \textit{approximate computing network-aware} and the \textit{network 
approximate-computing aware}?} If so, what design choices are required to effectively integrate approximate computing with production networks?''

Our answer to these questions is a network system called \emph{\sysname} that is designed for data-center approximate computing.
Behind \sysname, our insight is that existing reliable transports and queuing disciplines are \emph{conservative} and thus generally underutilize the network~\cite{Homa-Sigcomm18}; 
they aim to deliver \emph{all} packets while avoiding network loss. 
These design choices are suboptimal for approximate workloads which can tolerate data loss and can potentially run on a more aggressive network layer.  
We explore and propose two distinct points in the network design space for approximate computing: 
first, approximate transports which can send at a rate that slightly exceeds network capacity, 
and second, a network which can allocate significantly fewer queuing resources to approximate traffic.

Although these high-level design choices are simple, there are admittedly several critical challenges in realizing them in today's production networks in a manner that generalizes across the broad spectrum of approximate computing applications ranging from streaming and batch to machine learning.
First, although approximate applications can tolerate some slack, dropping data arbitrarily is not acceptable as these applications still need to meet certain accuracy requirements. 
Thus, \sysname\ needs to accurately infer the approximation slack and deliver optimal performance without violating accuracy requirements.
Second, an application's slack varies over time due to packet drops and network dynamics. Thus, \sysname\ needs to automatically and effectively adjust approximate traffic's behavior to react to evolving network 
dynamics and variability.
Third, the fundamental shift in approximate traffic's behavior towards more aggressive rates potentially incurs queuing delays and build-up.  Thus, \sysname\ must intelligently allocate network resources to approximate and non-approximate traffic
in a manner that is efficient and starvation-free while simultaneously addressing the spectrum of accuracy requirements for different approximate jobs. 
Finally, \sysname\ should offer an abstraction that is generic and expressive enough to support different types of approximate applications while being easy to use.


Intuitively, addressing the above challenging in an application-specific manner may be trivial. However, addressing them in an application-agnostic manner requires synthesizing a set of novel \emph{slack-based} network techniques across several layers of the network stack, including a slack-based transport layer, a slack-based switch queue design, and a slack-based abstraction and accompanying libraries.
At the core of \sysname\ is a dynamically adjusted, slack-based metric, {\em Target message Loss Rate}, or $TLR_{t}$, which captures an application's instantaneous slack tolerance at time $t$.
Crucially, $TLR_t$ evolves dynamically as the network state changes, with the goal that the final target loss rate is equal or slightly smaller than application's overall tolerable amount of loss. 

\sysname\ exposes a simple yet powerful interface to application frameworks such as Spark and PyTorch. 
To use \sysname, we include a developer-friendly user-level library, {\em \approxlib}, which acts as a bridge between application frameworks and our network system. In \S\ref{sec:framework}, we demonstrate \sysname's ease of use by porting four big-data application frameworks.

At the transport layer, we propose a new {\em Approximate Transmission Protocol}, or {\em \atp}, which adapts the application's sending rate to utilize all available network bandwidth opportunistically based on the instantaneous target loss rate. 
Internally, \atp\ further exploits approximate applications' tolerance of un-ordered data to minimize their job completion time with a new {\em approximate-aware} scheduling policy.

At the network layer, our \emph{slack-based network resource allocation} uses the {\em target loss rate} (and implied aggressiveness) to determine an application's priority and appropriately queues its traffic. To support this, 
\sysname\ configures lower-priority switch queues to a tiny queue size and associates approximate flows with more loss tolerance to these queues.
The highest priority queue and almost all switch buffer space are reserved for non-approximate traffic.
Together, these techniques ensure minimal retransmission, efficient bandwidth utilization, and fast job completion.
In particular, the switch settings improve the performance of non-approximate workloads while not affecting the performance or accuracy of approximate workloads.

We implemented \sysname\ as a user-level library at end hosts and by changing existing switches' configurations.
We ported four data-processing/ML frameworks, Kafka~\cite{apach-kafka}, Flink~\cite{apach-flink}, Spark~\cite{apach-spark}, and PyTorch~\cite{PyTorch}
to \sysname.
In addition, we implemented \sysname\ on the ns2 simulator~\cite{ns2-simulator} to understand our techniques under production scales.
In our simulation, we evaluated \sysname\ using large Fat-tree~\cite{Al-Fares-Sigcomm08} and 2-layer CLOS topologies
and two large-scale real-world traces~\cite{Atikoglu12-keyvalueanal, Roy-Sigcomm15}. 
We compared \sysname\ to UDP, DCTCP~\cite{Alizadeh10-SIGCOMM}, pFabric~\cite{Alizadeh-Sigcomm13}, Aeolous~\cite{Aeolus-sigcomm20}, and two sender-drop approaches, 
one that samples packets to be sent uniformly throughout a job and one that sends data as early as possible.
We evaluated our real implementation on a small cluster with two real-world workloads and three types of approximate computation.
Our evaluation results show that \sysname\ improves approximate application job completion time by up to 80\%.
Meanwhile, \sysname's measured loss rate is small 
and always below application-specified max loss rate.
Moreover, it achieves fairness across different (approximate and non-approximate) jobs and improves the performance of co-running non-approximate traffic by 79\%.

This paper makes the following contribution.

{
\begin{figure}[th]
\begin{minipage}{0.5\textwidth}
\begin{minipage}{0.48\columnwidth}
\begin{center}
\centerline{\includegraphics[width=0.9\columnwidth]{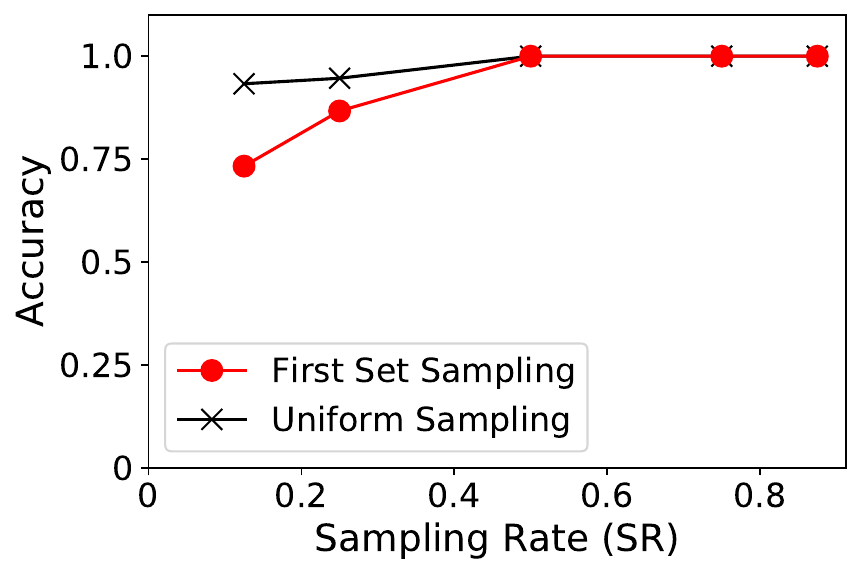}}
\end{center}
\vspace{-0.4in}
\mycaption{fig-motivation-top}{TopK Accuracy.}
{ 
}
\end{minipage}
\vspace{-0.2in}
\begin{minipage}{0.48\columnwidth}
\begin{center}
\centerline{\includegraphics[width=0.9\columnwidth]{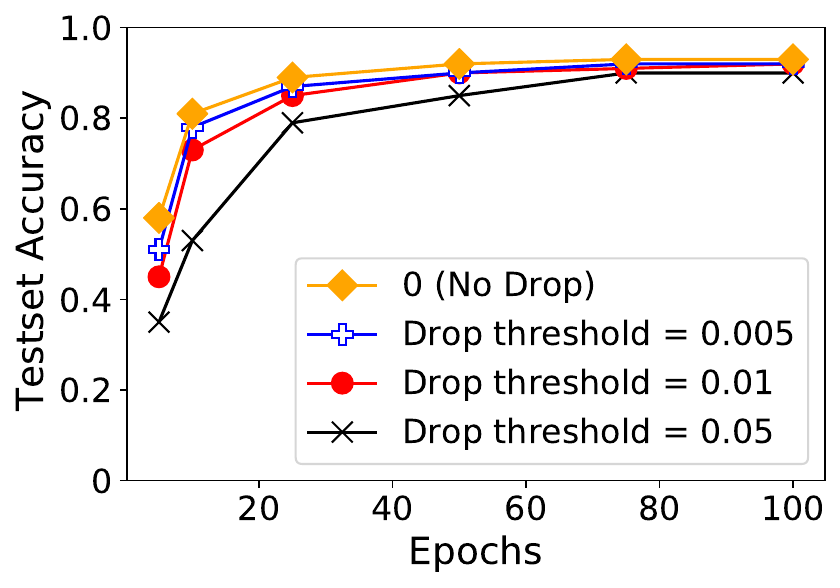}}
\end{center}
\vspace{-0.4in}
\mycaption{fig-motivation-ml}{Resnet50 Accuracy.}
{ 
}
\end{minipage}
\end{minipage}
\end{figure}

}

\vspace{-0.05in}
\begin{itemize}
\vspace{-0.05in}
\item Analysis of the limitations of existing approximation systems that are network oblivious.
\vspace{-0.05in}
\item The first proposal of incorporating approximate computing at the network layer. 
\vspace{-0.05in}
\item \atp, a new transport protocol designed for distributed approximate computing, and a set of approximate-aware switch resource allocation policies. 
\vspace{-0.05in}
\item Implementation of \sysname\ with both simulation and real implementation, and four real datacenter distributed frameworks ported to \sysname.
\vspace{-0.05in}
\end{itemize}
\vspace{-0.05in}

We will make our simulation and real implementation source code publicly available upon publication.

\section{Today's Approximate Computing}
\label{sec:background}

This section provides background on approximate computing and briefly describes the state of the art.

\subsection{Datacenter Approximate Computing}
\label{s:approxcomp}
Today, approximate computing paradigms in datacenters mainly target data analytics and machine learning applications 
running on batch and streaming platforms
where end-users do not need a precise answer and are perfectly content with an approximate answer. 
For example, an online service provider often needs to determine, in real-time, the top viewed webpages 
and runs a stream-processing job in their datacenter to do so.
This service provider does not care about the actual number of views, just their relative frequencies 

Figure~\ref{fig-motivation-top} shows the accuracy of running a Top-K (top 15) calculation on pick-up locations from the NYC taxi itinerary dataset~\cite{taxi-trace}). It can reach close to 100\% accuracy when the sampling rate is more than a half. Uniform sampling can further improve accuracy over a sampling scheme which only takes the first set of data points.

As another example, a large-scale machine-learning training job running on TensorFlow~\cite{TensorFlow1} or PyTorch~\cite{PyTorch}
can generate an accurate-enough model even with incomplete or imprecise training data or gradients.
We train the Resnet50 model~\cite{Resnet50} with the MNIST~\cite{mnist} dataset and drop gradients that are smaller than a threshold.
Figure~\ref{fig-motivation-ml} shows that approximate training has similar convergence rate as no approximate if the threshold is not too big, and the test-set accuracy is similar.


Although the designs of different approximate computing systems differ,
most of them follow a similar high-level work flow~\cite{ApproxJoin-SoCC18, StreamApprox2017, IncApprox-WWW16, Goiri-Asplos15, Park-SIGMOD18, Kandula-SIGMOD16}.
First, users specify their approximate workloads' requirements,
\eg, a maximum error rate. 
Then, the system chooses its own way of carrying out the approximation.
Finally, the system calculates or estimates the actual error from the execution and reports it to the user.

\subsection{Network-Oblivious Approximation} 
\label{sec:netoblivious}

Existing distributed approximate computing proposals~\cite{Agarwal13-EUROSYS,Ananthanarayanan14-NSDI,Goiri-Asplos15,StreamApprox2017} 
all build on a reliable transport (\eg, TCP) and treat the network as a black box.
They either sample data before sending them out or drop data after receiving them, both of which have their own problems.
Figure~\ref{f:aproxcomp:waterfall} illustrates the high-level flow of these approaches and \sysname.

\boldparagraph{Receiver-based sampling.}
This type of system sends all the data through the network,
and afterwards, the receiver discards a part of the data before processing it. 
The receiver could either drop some sampled data after all data is received (\emph{SampleRecv} Figure~\ref{f:aproxcomp:waterfall}(A))
or discard all data received beyond a threshold (\emph{DiscardRecv} Figure~\ref{f:aproxcomp:waterfall}(B)).
For example, StreamApprox~\cite{StreamApprox2017} improves Spark by dropping parts of the received input workload data  
and then start a Spark job to process the remaining data.
With receiver-based sampling, all data is transferred over the (reliable) network, 
resulting in significant resource waste in the network.
Moreover, with the scheme of Figure~\ref{f:aproxcomp:waterfall}(A), computation at the receiver will be delayed, impacting the entire application's performance.

{
\begin{figure}[t]
\begin{minipage}{\columnwidth}
\begin{center}
\centerline{\includegraphics[width=1.0\columnwidth]{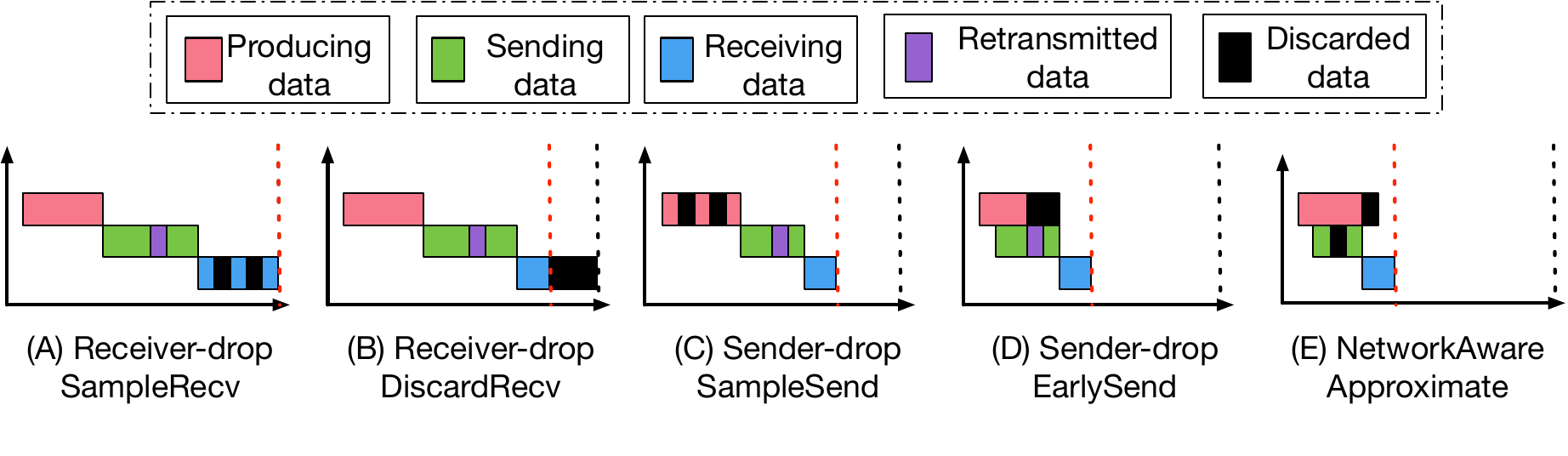}}
\vspace{-0.15in}
\mycaption{f:aproxcomp:waterfall}{Comparison of Different Approximate Computing Schemes.}
{
Waterfall of stages in data transfer.
Black line indicates the job completion time when no approximation is performed (baseline).
Red line indicates the job completion time of that scheme.
}
\end{center}
\end{minipage}
\vspace{-0.2in}
\end{figure}
}

\boldparagraph{Sender-based sampling.} 
Opposite to receiver-based sampling, sender-based sampling performs sampling at the sender 
before sending data out.
The sender could sample data as they are made ready (\eg, read from disk or received from another server) and send only the sampled ones, a scheme we call {\em \midsend} (Figure~\ref{f:aproxcomp:waterfall}(C)).
The sender could also send data as soon as they are ready 
and stop sending once enough data has been received, what we call {\em \earlysend} (Figure~\ref{f:aproxcomp:waterfall}(D)).
ApproxHadoop~\cite{Goiri-Asplos15} is an example of \earlysend.
It optimizes Hadoop's map phase by 
sending map results to reducers as soon as mappers generate them
and stops the mapper phase when user-specified accuracy is met. 
%

By sending only a subset of data, \midsend\ reduces network bandwidth consumption.
However, it is as slow as the speed of the entire data production process.
It misses opportunities to more aggressively use network bandwidth.
\earlysend\ improves application performance over \midsend\ by sending data as soon as they are ready.
However, by treating the network as a black box (\eg, using a reliable transport with a congestion-control algorithm that is approximate oblivious),
data could be sent at a conservative rate (because today's congestion control algorithms have the goal of minimizing packet loss) and/or unnecessary retransmission would be involved (because today's reliable transports deliver every packet).
Both these scenarios would delay the completion of application workloads.


\boldparagraph{Takeaway.} 
Both receiver- and sender-based approximate computing schemes are network oblivious,
resulting in excessive network bandwidth consumption, inefficient bandwidth consumption, floods of retransmission, and/or delayed job completion time.
We argue that network support for approximate computing can offer new opportunities in improving application performance 
and achieving better network resource allocation.
However, the lack of network awareness of today's approximate computing paradigms minimizes 
their ability to achieve these goals.

\section{\sysname\ Overview} 
\label{sec:overview}

{
\begin{figure}
\begin{center}
\centerline{\includegraphics[width=0.9\columnwidth]{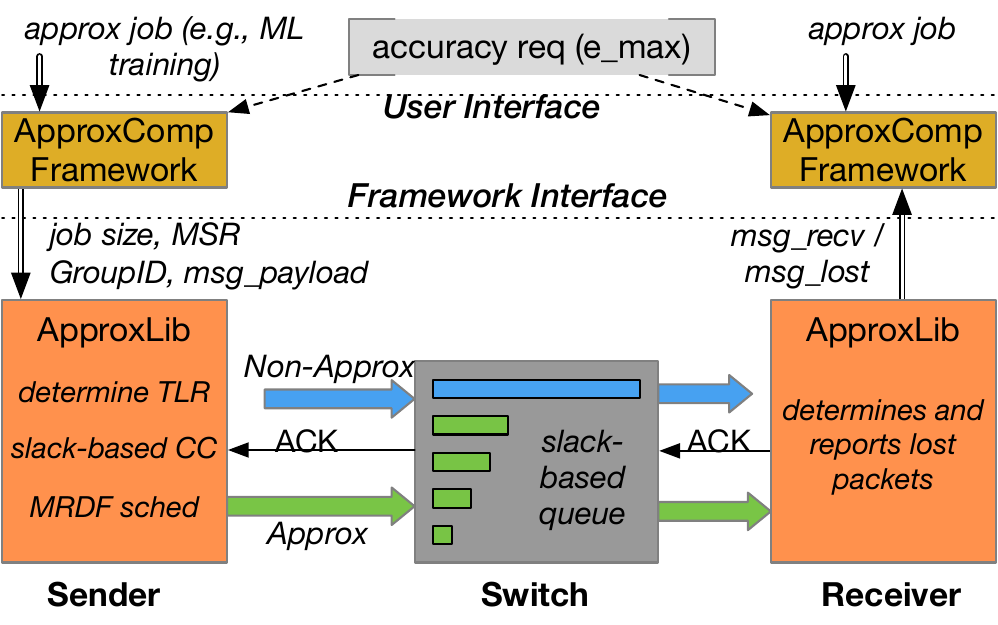}}
\vspace{-0.1in}
\mycaption{fig-overview}{\sysname\ Architecture.}
{
}
\vspace{-0.2in}
\end{center}
\end{figure}
}

This section presents an overview of our proposed framework, \sysname. 
The intuition behind \sysname's design is to use the slack allowed by approximate applications to determine
(1) an appropriate sending rate (by our new transport protocol),
(2) a policy to selectively retransmit lost packets (by our new packet scheduling policy), 
and (3) priority assignment and queue buffer space (by our new switch resource allocation mechanism). 


\subsection{Overall Architecture}
\sysname\ consists of several components (Figure~\ref{fig-overview}), all of which are approximate-aware and have \emph{slack-centric} designs. 
\emph{\atp}, the core of \sysname, is a new transport protocol that is neither unreliable (like UDP) nor completely reliable (like TCP).
Its core idea is to allow approximate traffic to send at rates that \emph{exceed network capacity} (in order to aggressively try completing approximate jobs sooner), but doing so in a controlled manner (so as to still meet application-specific accuracy requirements).

We perform auxiliary support for certain approximate applications' sampling requirements (\S\ref{sec:group-sampling}).
Together with \atp's end-host logic, they form a user-level library called \emph{\approxlib} running at end hosts.

At the network layer, to provide isolation and prevent starvation, we propose a new \emph{slack-based} switch queue configuration and scheduling algorithm (\S\ref{sec:switch-design}). 
The core idea is to categorize flows based on their \emph{current slack requirements} and to assign them to switch priority queues accordingly.

\subsection{\sysname\ Workflow}
To leverage \sysname, framework developers would port an approximate framework such as PyTorch and Spark to run on \approxlib.
End users use these ported frameworks in almost the same way as they do before.
Before a workload starts, an end user specifies the overall accuracy requirement of their job to the approximate framework, which translates it to workload approximation configurations for \approxlib.
At run time, the framework sends \emph{messages} to the network via \approxlib\ and may dynamically adjusts the approximation configurations.
\approxlib\ determines the dynamic target loss rate, packet sending rate, sending order, and retransmission criteria based on the approximation configurations given by the framework and the current network congestion.
At the receiver side, \approxlib\ reports lost messages to the framework, which is used to adjust new sampling criteria.
After a job completes, the framework reports the application-level accuracy, profiling statistics such as the measured loss rate, and the computation results to the user.
\subsection{\sysname\ Interfaces}
\label{sec:approx-support}

\approxlib provides two distinct interfaces; one for end users, to specify acceptable parameters for their specific job, and one for the approximate framework developers to integrate \sysname\ into their framework.  A fundamental challenge in designing the framework abstraction and interface lies in ensuring that the interface is simple but powerful enough to support the broad set of approximate applications which exist today. In particular, while batch and streaming often adhere to similar principles, deep learning explores a different point in the design space. 



\subsubsection{User Interface} 
Users carry out approximate computation on top of ported frameworks by first defining certain accuracy requirements: 
the maximum acceptable error rate for a job ($e_{max}$).
Error bounds are applicable to almost all approximate applications such as ML, batch and streaming data-processing systems.
When a user specifies $e_{max}$, then \sysname\ tries to finish the job as soon as possible while ensuring that the error is within $e_{max}$. 
Afterwards, users submit {\em job}s to the ported frameworks in the same way they would on the original framework.
For example, a job can be a word count Spark job, one minibatch of machine-learning training task executed by PyTorch, or the processing of a window of streaming data.

\subsubsection{Framework Interface}  
\label{sec:frameinterface}
\boldparagraph{Per-job description.}
The approximate application framework ported to \approxlib\ first informs \approxlib\ about the size of a job $N$ before starting the job.
$N$ can be the total dataset size for batch systems or window size for streaming systems, both of which are known by the corresponding frameworks.
The framework is also responsible for translating user-specified accuracy requirements (\ie, $e_{max}$) to a minimal message sampling rate (\msr) that will be passed to \approxlib.
Different frameworks would have their own ways of doing this translation.
A common practice we suggest is to keep estimating the potential error based on the received and the dropped data, compare it with the user-specified $e_{max}$, and adjust \msr\ in a way that would minimize the error difference.
We elaborate in \S~\ref{sec:framework}.
%


\boldparagraph{Per-message description.}
At run time, when the framework executes a {\em job},
it sends multiple {\em messages} to \approxlib, each of which in turn includes one or more packets.
Only when all the packets in a message are successfully delivered does \sysname\ treats the message as being delivered,
and \sysname\ guarantees to deliver at least $MSR \times\ N$ messages for a job.
Frameworks that desire fine-grained sampling could make one data unit as an individual message (with the tradeoff of potential performance overhead).
Frameworks running on \approxlib\ can use unmodified POSIX socket APIs 
to send and receive messages in a job.
In the beginning of the socket API payload, we leave a header field for frameworks to specify per-message approximate information, which we will describe next. 

\subsubsection{Expressing Sampling Criteria}
\label{sec:group-sampling}
By default, \approxlib\ treats every message in the same way, \ie, \approxlib\ is free to choose any messages to drop.
However, many approximate computing frameworks have specific requirements on how data should be sampled, 
and for these frameworks, not all messages can be treated in the same way. 
For example, frameworks like BlinkDB~\cite{Agarwal13-EUROSYS} 
perform queries using \emph{stratified sampling},
where data is separated into different subsets of data 
(\eg, based on certain database fields)
and samples are drawn with an equal probability within each subset.
Other frameworks could choose \emph{quota sampling}, where the sampled data size (not probability) is the same across subsets of data.
Yet, some frameworks may desire to precisely control what data can and cannot be dropped,
\eg, a machine-learning framework may want to ensure that all messages carrying large gradients are delivered and is OK with the loss of small gradients.

We introduce a concept of {\em group} to support a broad range of sampling patterns and requirements effectively.
To use this construct, the framework specifies a \emph{GroupID} in the header of each message,
and for different \emph{groups}, the framework could specify different \msr{}s (\eg, 0.8 for GroupID 1, 0.5 for GroupID 2, 1 for GroupID 3).
\sysname\ separates messages into groups based on their GroupID and guarantees to deliver the minimal samples within each group based on their \msr{}s.
If \msr\ is 1, then \sysname\ will treat the group as a non-approximate workload and delivers every message in order.
The ability to employ different \msr{}s for different group allows \sysname\ to implement a broad range of sampling techniques.
For example, to express stratified sampling, a framework could separates its dataset into different groups and use the same \msr\ for each group,
while for quota sampling, a framework would use different \msr{}s to achieve the same sampled data size across groups.
Frameworks could also put data that cannot be lost (\eg, certain metadata) into a group with \msr\ 1.
If no GroupID is specified, \approxlib\ will treat all messages as part of one global group that uses the global \msr\ specified by the framework.

\section{Approximate Transport Protocol (\atp)}
\label{sec:atp}
Here we discuss the principles and design choices underlying our Approximate Transport Protocol (\atp). Our core contribution in designing \atp\ is to introduce a method for enhancing existing rate-control algorithms with the notion of slack. Specifically, while existing algorithms focus on detecting and reacting to network state, we provide principles for showing how to integrate the application's approximate nature -- thus allowing rate-control and packet scheduling to react both to network loss and application's inherent slack -- and illustrate these principles by enhancing DCTCP.

Our design for \atp\ centers around several principles: first, our protocol should seamlessly co-exist with other transport protocols, e.g., DCTCP; second, to easily interoperate with modern approximate applications (e.g., streaming, batch, and ML frameworks), it should be message-oriented with each message consisting of one or more packets\footnote{Traditionally, each packet's header includes metadata required by the receiver to reconstruct the message, \eg, \seq\ and \msgid.}; and, third, our protocol should support jobs with varying levels of slack (even those that can not tolerate any loss).

To optimize application job completion time while preserving accuracy requirements (\ie, successfully send \msr\ amount of a job's data as specified by the approximate framework), 
an ideal approximate transport should send data at an \emph{aggressive a rate as possible} -- essentially in a rate that results in the network dropping at most the number of packets that the application can tolerate, i.e.,  (\ie, $(1-MSR) \times N$). 

This differs distinctly from traditional transports, which attempt to prevent packet loss. The core fact that approximate transport embraces loss significantly transforms and simplifies congestion control. First, unlike existing congestion control which includes complex mechanisms for loss recovery, e.g., timeout and dupACKs, an approximate transport does not need to include these. Instead, the transport comprises of a simple but more aggressive rate control algorithm whose aggressive arises because it reacts to the application's approximate nature in additional to network state (\S~\ref{sec:rate-control}). Second, the fact that in-order delivery is not crucial implies that the protocol can also eliminate the mechanism required to pace and ensure in-order delivery (\S~\ref{sec:mrdf}).

\subsection{Strawman Protocol}
\label{sec:atp-strawman}
Based on this intuition, our strawman protocol, what we call \emph{\atpfixtlr}, determines the \emph{target message loss rate}, or \emph{\tlr}, for a job to be $1-\msr$ (specified by framework). Given this \emph{\tlr}, \atpfixtlr\ alters the rate control of an existing transport to adjust sending rate in response to \tlr instead of traditional signal of congestion (e.g., packet loss). Specifically, it then measures the rate of actual data loss and increases (decreases) the sending rate if the measured loss rate is lower (higher) than \tlr. Moreover, if the measured loss rate is still too high after decreasing the sending rate, it would retransmit lost packets. This minor change to rate control enables an existing transport to be approximate aware.
 
The key issue with this strawman approach is that \tlr\ is statically defined; however,  in production networks, there is no way to predict how much data the network will drop perfectly. Additionally, fluctuations in background traffic will also lead to excessive or insufficient loss.
As a result, the application's slack tolerance changes over time as more packets are delivered, or more packets are lost. 
Thus, we need to dynamically adjust \tlr\ to capture the application's \emph{instantaneous} slack.

\subsection{Dynamic Slack-Based Rate Control}
\label{sec:rate-control} 
We propose a novel {\em approximate-aware, slack-based} rate control algorithm that simultaneously adapts to network states, controls the magnitude of packet loss, and ensures efficient network resource utilization for both approximate and non-approximate workloads.
Unlike existing congestion-control algorithms that use network signals (\ie, ECN, loss, or RTT) as the sole input for their {\em congestion control algorithm} (approximation-unaware),
\atp\ uses workload-specific (instantaneous) slack-information in addition to network signals to inform its congestion control algorithm.

Our congestion-control intelligence revolves around the notion of a dynamically-adjusted \tlr.
Intuitively, $TLR_t$ at time $t$ should account for how actual loss rate in the past has deviated from \tlr{}s at those times.
However, it is difficult to incorporate all past deviations in an equation to deduct a new \tlr.
Instead, we could approach the problem from a different angle: 
how much data could potentially be lost in the future?
With this idea, we have the following equation, which determines $TLR_{i+1}$ at the end of every \emph{epoch}\footnote{By default, an epoch is one RTT in our algorithm.} $i$:

\vspace{-0.2in}
\begin{equation}
\label{eq:tlr}
    TLR_{i+1} = \frac{(1 - MSR) \times\ N - TotalNumLostMsg}{N - TotalNumDeliveredMsg}
\end{equation}
\vspace{-0.2in}

where $N$ is the total number of messages in a job 
and $(1 - MSR) \times\ N$ is the maximum number of messages that could be lost for the entire job.
The numerator of this equation is thus the number of messages that can still be lost in the future.
The denominator is the total number of messages that could potentially be sent, which include both unacknowledged messages and new (\ie, future) messages that have never been sent out before.
Thus, $TLR_{i+1}$ is the potential message loss rate that the job could tolerate in the future.
Note that with this equation, $TLR_0$ is initialized to $1-MSR$, the same as the static value used in \atpfixtlr.

\textbf{Enhancing DCTCP:} Recall that existing rate-control algorithms for today's transport protocol, e.g., DCTCP, adjust the sending rate by comparing the current loss against a predefined threshold. Instead, with \atp, current loss is compared against a dynamic target threshold, essentially $TLR_{i}$. %
%
More specifically, \atp\ continuously monitors the current loss rate, $\rho_i$~\footnote{ by comparing the ACKs received and the total number of packets sent in an epoch $i$, with $\rho_i = 1 - \frac{N_{ack}^i}{N_{sent}^i}$}.
If $\rho_i$ is less than $TLR_i$, then \atp\ increases its sending rate:

\vspace{-0.2in}
\begin{equation}
R_{i+1}=(1-m)\times R_i+m\times{R_{max}}
\end{equation}
\vspace{-0.2in}

where $m$ is a parameter which controls \atp's reaction speed, and it trades off between convergence speed and network utilization.
$R_{max}$ is the highest rate that the host can send at, \ie, the NIC port capacity.

On the other hand, if the current loss rate, $\rho_i$, is higher than the $TLR_i$, then \atp\ decreases its sending rate:

\vspace{-0.1in}
\begin{equation}
 R_{i+1}=R_i\times(1-\frac{\rho_i}{2})
\end{equation}
\vspace{-0.2in}

Note that the above rate adjusting algorithms follows DCTCP's (although the criteria for invoking these algorithms are different and are approximate aware). Our choice of DCTCP over other contemporary algorithms stems from its proven stability and convergence properties. Also DCTCP fact that it does not explicitly focus on short flows.


\textbf{Preventing Starvation:} There is one caveat that needs to be taken in the above algorithm.
In a highly congested network, ACKs may not be delivered promptly.
Without getting ACKs, \atp\ would decrease both the sending rate (equation 3) and $TLR$ (equation 1);
the latter would further cause the sending rate to decrease.
With a sending rate that is too small, \atp\ will not be able to detect or react to network state, causing a starvation problem.
To prevent this starvation case, \atp\ mandates a minimum sending rate ($R_{min}$), which allows flows to probe the network periodically.
We set $R_{min}$ to 1 packet per RTT.

\subsection{Approximate-Aware Packet Scheduling}
\label{sec:mrdf}
Apart from the tolerance to lost data, most\footnote{Certain approximate applications like live video analytics~\cite{Haoyu-NSDI17, Sen-Sigcomm10} have dependencies across messages. They can disable our MRDF scheduling. }
approximate applications can also tolerate re-ordering of data,
\eg, $Avg(A,B,C)$ is the same as $Avg(C,A,B)$.
We exploit this feature by introducing a new scheduling policy called {\em minimal-remaining-data-first} (MRDF), 
which exploits the opportunity of approximate applications' tolerance of unordered data to send messages out of order for better job completion time.
   Specifically, an \atp\ sender (its \approxlib) calculates the amount of data that is left to be sent for each message (\ie, the total size of the message minus the size of data successfully delivered to the receiver).  Given information about message sizes, \atp\ then sends packets using the shortest remaining data policy.
Note that in this process \atp\ can choose both un-acknowledged packets and new packets that have been never sent out before.
Unlike other possible transports that either always re-transmit lost packets or always wait for new packets to arrive, our MRDF scheduling mechanism could achieve the best performance. 

\section{Slack-Centric Network Resource Allocation}
\label{sec:switch-design}

Fundamentally, approximate applications place different demands on the network than traditional applications: they require fewer network resources (because of their loss tolerance). However, if not controlled, they could end up \emph{unfairly} consuming too many network resources (because of their aggressive sending rates).
These features motivate us to rethink switch resource allocation mechanisms. 


\subsection{Strawman Approach}
\label{sec:switch-strawman}
Approximate workloads could work with small network resources (\eg, lower switch priority queue, less switch queue buffer), but its more aggressive sending rate would unfairly impact non-approximate traffic.
Based on this intuitive idea, our strawman approach separates approximate and non-approximate traffic into two switch priority queues.
It devotes the highest priority and most (or even all) switch buffer space to non-approximate traffic so as to improve the performance of non-approximate workloads.
%
This seemingly feasible approach has a key issue: 
all approximate traffic share the same queue, but not all approximate traffic have the same slack tolerance.
An approximate application that tolerates more loss could potentially send its traffic more aggressively than one that tolerates little loss and starve the latter.

An improved solution (what we call \textit{\atpfixqueue}) is to use multiple queues to differentiate approximate workloads with different slack tolerance
and statically assign a job to a priority queue based on its \msr.
However, as network states and application states change, a job's tolerance to loss is dynamic, as discussed in \S\ref{sec:rate-control}.
Thus, we need a mechanism that could adjust a job's switch queue assignment dynamically.


\boldparagraph{Insight.}
Our insight is that it is beneficial to: (1) allocate more buffer space and assign higher priority to applications that can tolerate less or zero slack, 
(2) assign applications with similar slack to the same priority to ensure that only application with similar level aggressiveness compete with each other, 
and (3) dynamically adjust an application's priority based on its instantaneous loss tolerance.

\subsection{Dynamic Priority and Queue Allocation}
\label{sec:dynamic-network-alloc}
We propose a dynamic mechanism that assigns different network resources based on a job's \emph{instantaneous} slack tolerance.
Below, we discuss the detailed design.

\boldparagraph{Slack-based priority assignment.}
First, like the strawman approach, we reserve the highest-priority queue only for non-approximate jobs.
We use the remaining switch queues\footnote{Existing commodity switching chips supports multiple queues 
(typically 4 to 8~\cite{PIAS-nsdi15, Wei-nsdi16, Bai-CoNext16}) per egress port.} for different approximate traffic and associate different 
\textit{\tlr thresholds} to each of them, \ie, $ThreshTLR_k$ for the $k$th priority queue, with $k=0$ being the highest priority queue.
Intuitively, lower priority queues could result in more packet loss and should be used for jobs with more slack tolerance. 
Thus, $ThreshTLR_k$ should be higher for larger $k$.
We empirically evaluated different approaches assigning thresholds to the queue using the data center traces and workloads described in \S~\ref{sec:simulation}.  We observed that a simple heuristic which evenly distributes threshold values across 0 and 1 performed best and this was comparable to exponential assignment used in prior works.
For example, with 8 switch queues, we have $ThreshTLR_0 = 0$, $ThreshTLR_1 = 0.125$, ..., $ThreshTLR_7 = 0.875$.

Instead of statically assigning jobs to switch queues, we dynamically direct a job's traffic to an appropriate queue based on its current \tlr. 
Specifically, for each epoch, $i$, we direct flows with $TLR_i$ that is greater than $ThreshTLR_{k-1}$ but no smaller than $ThreshTLR_{k}$ to the $k$th queue.





\boldparagraph{Slack-based queue buffer space allocation.}
To determine the buffer size for the different queues, we empirically evaluated different allocation strategies and found the following simple scheme that works well across workloads.  We set the switch's lowest priority queue to use a 1-MTU buffer size and increase the buffer size by 1 MTU for every higher priority 
(\ie, the $k$th lowest priority queue size is $k$ MTU), until the second highest priority queue.
Then, we leave all the remaining switch buffer to the highest priority queue (non-approximate messages).

A smaller switch queue size for lower-priority queues would cause more data loss,
which is acceptable for flows with larger $TLR_i$ because they can tolerate more loss.
On the other hand, if we were to assign a larger queue to these flows, 
they would fill up the queues and increase the queue delay, 
causing packet loss signals to propagate slowly to the receiver and back to the sender.
This delayed signal further prevents the \atp\ rate control from quickly making the right decision, causing more packet loss and retransmission.
Lastly, a large switch buffer allocation for approximate flows underestimates loss rate estimation, 
as it saves packets, which makes \atp\ overestimates its sending rate. 
By using small queues for approximate traffic, 
switches can save most of its buffer space for \nonapproximate\ traffic.
Doing so improves \nonapproximate\ traffic's performance (\ie, reduces delay) 
and/or enables switches to handle more \nonapproximate\ traffic. 

 Allocating buffers to queues and assigning thresholds are currently empirically derived parameters. We believe that these parameters will need to be re-evaluated in drastically different topologies or workloads. As part of future work, we plan to employ existing system approaches that use machine learning to tune system configurations. 

\section{Porting Frameworks to \sysname}
\label{sec:framework}

To demonstrate \sysname's ease of use and 
to evaluate \sysname\ with real applications, 
we ported four popular datacenter big data frameworks to \sysname: one batch (Spark~\cite{apach-spark}), two  streaming systems (Kafka~\cite{apach-kafka}, Flink~\cite{apach-flink}) and one ML system (PyTorch~\cite{PyTorch}). Below we elaborate on how these frameworks leverage \approxlib interface.

\boldparagraph{Batch data-processing frameworks.} 
Batch data-processing frameworks like Spark processes data in batches which are large sets of data of whose total size is known apriori~\cite{Vojislav-NSDI19}.
To port batch frameworks to \approxlib, their developers modify the framework to first specify the total job size (i.e., batch size) to \approxlib.
Then, at runtime, the user specifies an acceptable error, $e_{max}$, when a job is submitted and the framework translates this into \msr\ which is also pass to \approxlib\ along size the job size. 
While different computations require different amount of data to reach $e_{max}$, we can deduce a good sampling rate based on sampling theories for most computations like \avg\ and \rank\ in the following way.
We first set the initial \msr\ based on heuristics \eg, \msr=0.9 for $e_{max}=10\%$.
As more data is received, we use well understood sampling theory~\cite{Sampling, Sampling-1}, similar to prior work on approximation~\cite{IncApprox-WWW16, Goiri-Asplos15, ApproxJoin-SoCC18}, to estimate the error based on the mean and variance of the received data. We then adjust \msr\ to make the error closer to $e_{max}$.
\approxlib\ determines \tlr\ by applying this adjusted \msr\ in Equation~\ref{eq:tlr}.

Batch frameworks often employ grouping (\eg, grouping on a data field). \sysname\ supports grouping and allows the user to specify acceptable error for individual groups. At the framework level, developers associate each message with a GroupID.
\sysname\ optimizes the job completion time while meeting this group-based sampling specification.

We ported Spark to \sysname\ by performing approximation at two stages:
1) the data input stage,
\ie, network communication when Spark workers read input data,
and 2) the intermediate stage where mappers send data to reducers.
Optionally, \sysname\ allows users to enable approximation at either or both stages.
In total, porting Spark took 320 lines of code and 3 engineering days.

\boldparagraph{Streaming data-processing frameworks.} 
Unlike batch frameworks, streaming frameworks like Kafka and Flink process data 
over rolling windows (\eg, time-based or count-based).
At the end of a window, streaming frameworks usually perform some computation 
of the data in the window, which often allows some approximation,
\eg, counting the number of web page views per minute.
A \sysname\ job for streaming frameworks would thus be data to be sent in a window.
Streaming frameworks specify the number of messages in a window (\ie, job size)
to \approxlib, together with $e_{max}$.
For streaming frameworks that work with time-based windows, they could estimate the total size of the data in a window based on statistics collected from previously windows (\eg, data arrival rate).

We ported Kafka and Flink by performing approximation 
at the stages where Kafka feeds data to Flink workers and where Flink performs the flat-map and reduce-by-key operations.
In total, porting these frameworks took 473 lines of code and 3 engineering days.

\boldparagraph{Machine-learning frameworks.}
Large-scale machine-learning training usually involves a distributed set of servers, each equipped with one or more GPUs~\cite{PS-sosp19, Tiresias-NSDI19, BytePS-osdi20}.
A significant performance overhead in distributed ML training frameworks is the communication cost across servers~\cite{Pipedream-sosp19, DDL-model-eval}.
Specifically, one of the most costly communication tasks is the \emph{allreduce} step in data-parallel ML training frameworks~\cite{TensorFlow, PyTorch},
which involves different servers exchanging their locally computed gradients.
However, not all gradients are of the same importance:
smaller gradients (whose values are close or equal to zero) are less important than big gradients.
Thus, smaller gradients could potentially be dropped with little or no impact on training accuracy~\cite{Casync-sosp21, lin2018deep, TernGrad}.

{
\begin{figure*}[th]
\begin{minipage}{0.245\textwidth}
\begin{center}
\centerline{\includegraphics[width=1.0\columnwidth]{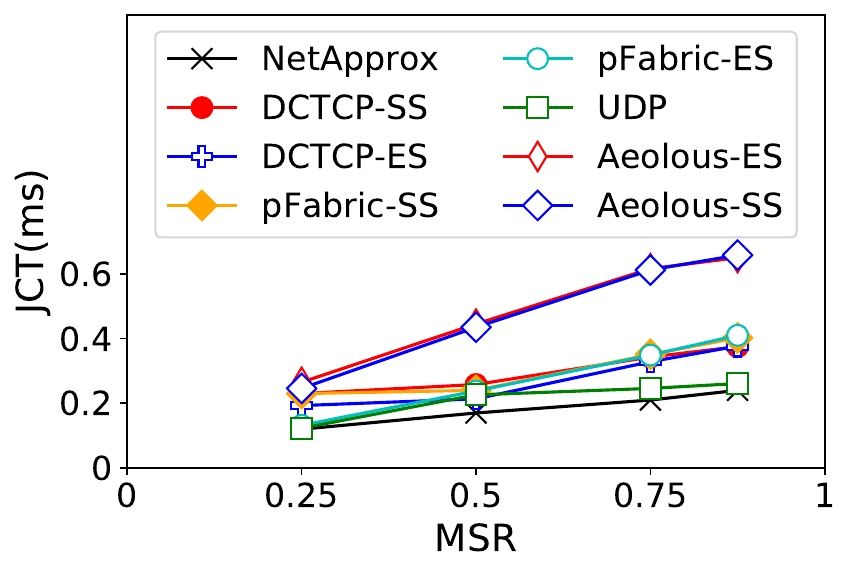}}
\vspace{-0.1in}
\mycaption{fig-jct-fbkv}{JCT of FBKV.}
{
}
\end{center}
\end{minipage}
\begin{minipage}{0.245\textwidth}
\begin{center}
\centerline{\includegraphics[width=1.0\columnwidth]{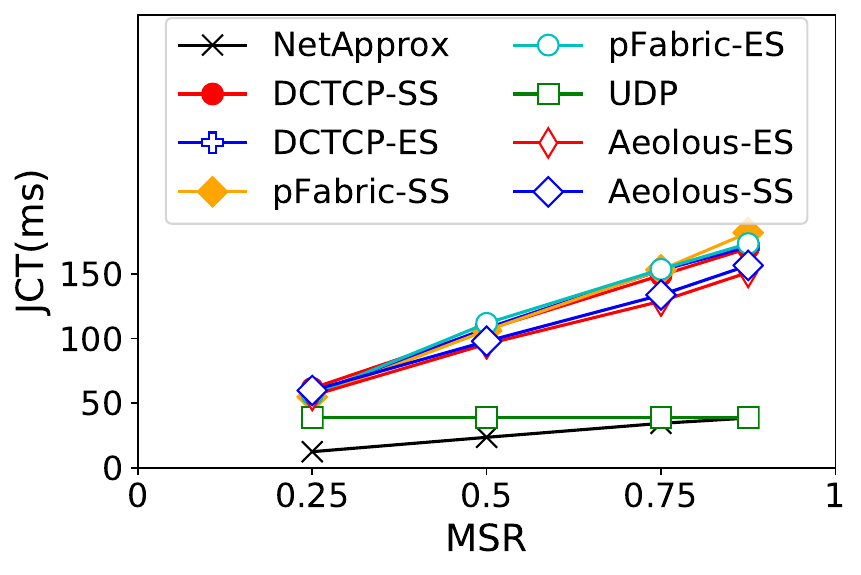}}
\vspace{-0.1in}
\mycaption{fig-jct-fbhd}{JCT of FBHD.}
{
}
\end{center}
\end{minipage}
\begin{minipage}{0.245\textwidth}
\begin{center}
\centerline{\includegraphics[width=1.0\columnwidth]{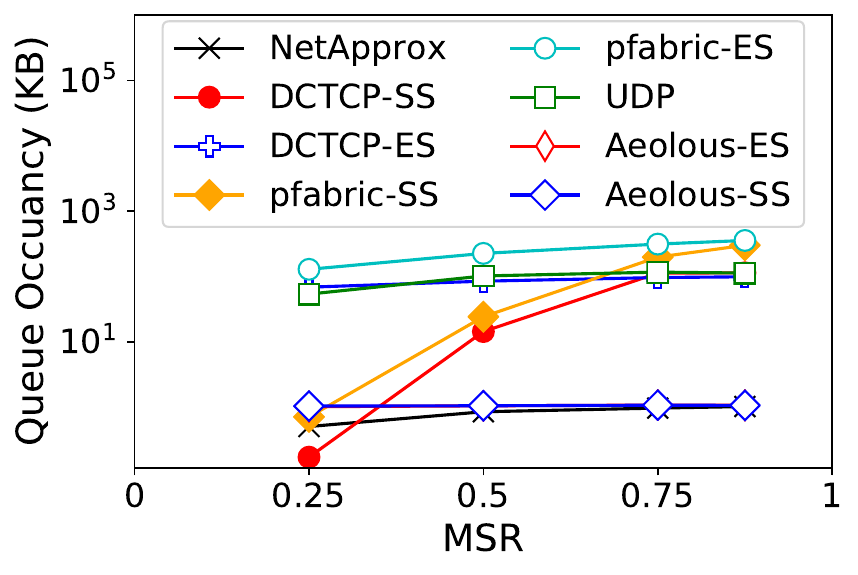}}
\vspace{-0.1in}
\mycaption{fig-queue-fbkv}{Queue Length of FBKV.}
{
}
\end{center}
\end{minipage}
\begin{minipage}{0.245\textwidth}
\begin{center}
\centerline{\includegraphics[width=1.0\columnwidth]{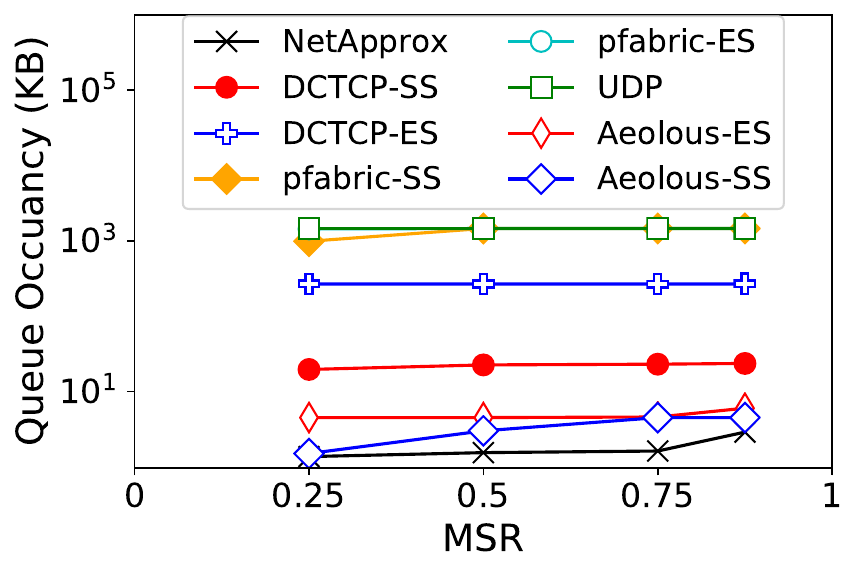}}
\vspace{-0.1in}
\mycaption{fig-queue-fbhd}{Queue Length of FBHD.}
{
}
\end{center}
\end{minipage}
\vspace{-0.2in}
\end{figure*}
}

We ported PyTorch to \sysname\ by changing the \emph{gloo ring-allreduce} functionality.
We use different groups to categorize gradients (\eg, a group of gradients that are smaller than 0.001, a group in between 0.001 and 0.0025, and a group that is larger than 0.0025) and use different \msr{}s for them.
The PyTorch \emph{ring-allreduce} groups gradients into \emph{segments} and then sends a segment at a time over the network.
When \emph{all} the gradients in a segment fall within the threshold of a group, we mark the segment to use the corresponding \msr\ of the group,
When a receiver is notified (by \approxlib) of a lost segment, 
it will use zero as the value for all the gradients in it.
In total, porting PyTorch took us 811 lines of code and eight engineering days.

With our ported PyTorch, instead of directly generating grouping criteria (gradient thresholds and \msr\ for each group) from a user-specified accuracy target which is extremely hard if not impossible, we view them as hyper-parameters that can be tuned in a similar way as traditional hyper-parameters like mini-batch size and learning rate.

\section{Evaluation}
\label{sec:results}

We evaluate \sysname\ using a combination of large scale simulations and testbed experiments which allows us to understand how our system works at a large scale and in practical settings with popular approximate applications.

\boldparagraph{Implementation details.} 
We implemented \approxlib\ 
as a user-space library in 1554 lines of C++ code.  
Applications can dynamically link \approxlib\ using \preload\, which intercepts POSIX system calls. 
\approxlib\ supports common POSIX socket APIs 
like \texttt{connect}, \texttt{accept}, \texttt{close},
\texttt{write}, \texttt{read}, \texttt{writev}, \texttt{sendfile64}, 
\texttt{send}, \texttt{sendmsg}, and \texttt{recv}.
Our current implementation of \approxlib\ transparently intercepts network system calls 
and call corresponding ATP network APIs.
We implement ATP as a user-space network stack using DPDK in 5386 lines of C++ code.

\subsection{Simulation Results}
\label{sec:simulation}
\boldparagraph{Simulation details.} 
For our simulator, we extended ns-2~\cite{ns2-simulator} to include \sysname's end-host and switch functionalities. 
Our simulations are performed on two common datacenter topologies; 
a k=12 fat-tree topology~\cite{Al-Fares-Sigcomm08} with 192 hosts
and a traditional two-tier Clos topology~\cite{PIAS-nsdi15, Wei-nsdi16, Bai-CoNext16} with 144 hosts.
For each topology, we evaluated our system with 40\gbps\ and 100\gbps\ line rate.
Because of space constraints, we only present 40\gbps\ FatTree results in the paper.
The other topologies' results are all similar qualitatively.

We set the switches to use eight priority queues~\cite{PIAS-nsdi15, Wei-nsdi16, Bai-CoNext16} 
with a total buffer space of 1.5\MB~\cite{Cho17-SIGCOMM}.
\sysname\ configures the eight queues in an approximation-aware way as described in \S~\ref{sec:switch-design}.
%
The \nonapproximate\ queue applies the ECN marking scheme with a 
marking threshold of 65~\cite{Alizadeh10-SIGCOMM, Cho17-SIGCOMM}, while the approximate queues starts to drop packets if the queues are full. 

We compare \sysname\ with two sender-side approximation schemes on three reliable transports and simple UDP (which provides a performance baseline).
The two sender-side approximation schemes are \earlysend\ (ES) and \midsend\ (SS).
ES sends messages as early as possible over a reliable transport and stops sending once MSR data have been received.
SS samples messages uniformly as they are ready (initiated by the application) with a sampling rate of MSR; it sends out the sampled message over a reliable transport.
The three reliable transports are DCTCP~\cite{Alizadeh10-SIGCOMM}, a widely used data-center transport that performs ECN-based adaptive congestion control;
pFabric~\cite{Alizadeh-Sigcomm13}, a priority-based transport that schedules packets based on their priority and performs simple rate control;
and Aeolous~\cite{Aeolus-sigcomm20}, a recently proposed transport aimed for high-speed network which initially sends BDP amount of packets at the line rate and then switches to a credit-based rate control algorithm~\cite{Cho17-SIGCOMM} afterwards. For Aeolous and pFabric, we reuse their opensource codebases.

\boldparagraph{Simulation workloads.}
We simulate two workloads using the
Facebook key-value-store trace (\FBKV{})~\cite{Atikoglu12-keyvalueanal} and the Facebook Hadoop trace (\FBHD{})~\cite{Roy-Sigcomm15}.   
We use the distributions of these traces to determine the size and inter-arrival times of messages in the jobs that we simulate.  
These two traces represent two extreme points in the spectrum of workloads: 
at one end \FBKV\ consists of jobs with small message sizes and number of messages, 
whereas \FBHD\ provides the other end of the spectrum with larger jobs.

\subsubsection{Overall Performance}
\label{sec:appperf}

{
\begin{figure*}[th]
\begin{minipage}{0.245\textwidth}
\begin{center}
\centerline{\includegraphics[width=1.0\columnwidth]{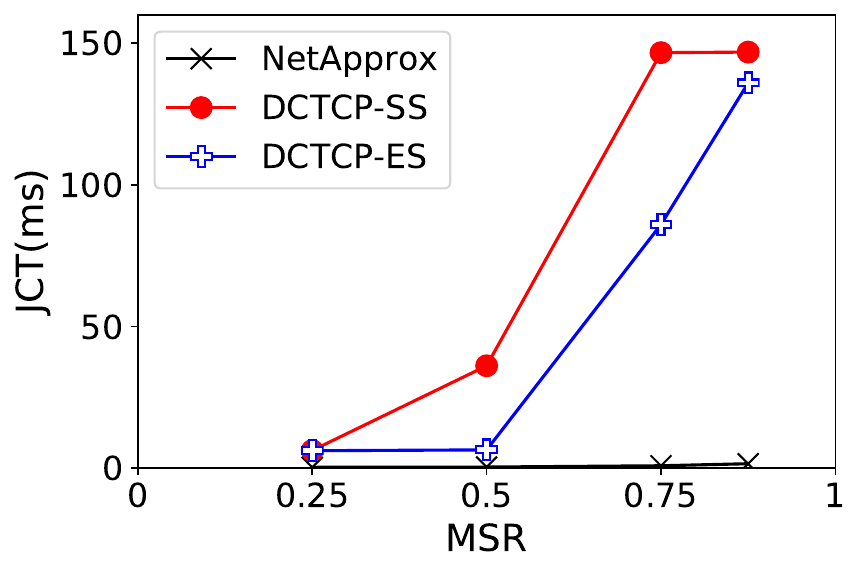}}
\vspace{-0.1in}
\mycaption{fig-all-to-one}{JCT of All-to-One Traffic.}
{
}
\end{center}
\end{minipage}
\begin{minipage}{0.245\textwidth}
\begin{center}
\centerline{\includegraphics[width=1.0\columnwidth]{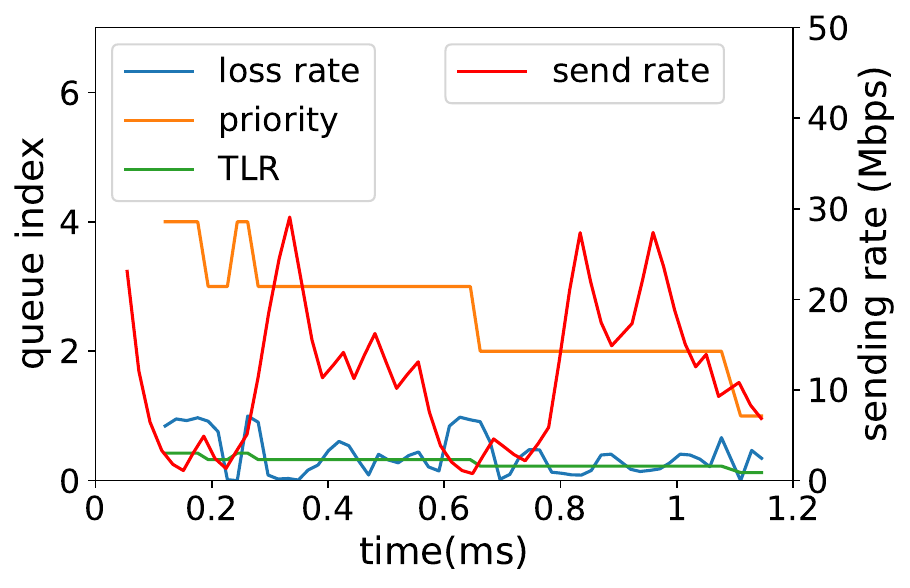}}
\vspace{-0.1in}
\mycaption{fig-dynamic-timeline}{Timeline of Dynamic Mechanisms.}
{
}
\end{center}
\end{minipage}
\begin{minipage}{0.245\textwidth}
\begin{center}
\centerline{\includegraphics[width=1.0\columnwidth]{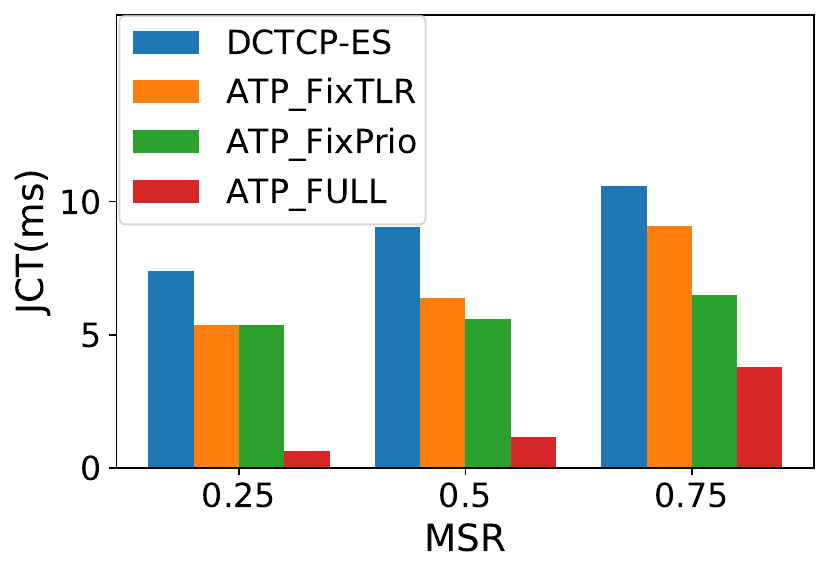}}
\vspace{-0.1in}
\mycaption{fig-dynamic-overall}{Effect of Dynamic Mechanisms.}
{
}
\end{center}
\end{minipage}
\begin{minipage}{0.245\textwidth}
\begin{center}
\centerline{\includegraphics[width=1.0\columnwidth]{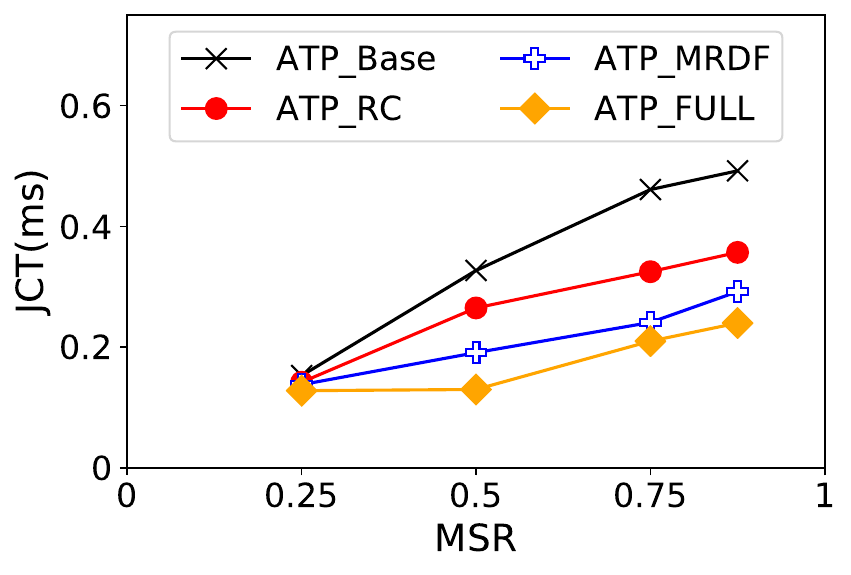}}
\vspace{-0.1in}
\mycaption{fig-atp-techniques}{Effect of \atp\ Techniques.}
{
}
\end{center}
\end{minipage}
\vspace{-0.2in}
\end{figure*}
}

We begin by evaluating the general performance of \sysname\ (Figures~\ref{fig-jct-fbkv} and \ref{fig-jct-fbhd}). 
For these experiments, we use an all-to-all traffic pattern (\ie, all hosts in the network are senders, each randomly selecting a host as the receiver).
Unsurprisingly, \sysname\ outperforms all reliable-transport-based techniques for both workloads. 
\sysname\ outperforms \midsend-based techniques because they drop packets at the sender at a constant rate 
even when the network has sufficient bandwidth to support more approximate traffic.
\sysname's sending rate is adaptive to the network's status, resulting in better network utilization and shorter \jct. 

\sysname\ also outperforms \earlysend-based schemes because \earlysend\ sends data aggressively in the initial period, which could cause congestion and performance overhead of retransmission. 
Moreover, \sysname\ saves more network resources, since it actively works to ensure that these flows use less than their network fair share, whereas pFabric, DCTCP, and Aeolous try to ensure that they use their ``fair'' share of network bandwidth.
Aeolous has the worst performance for \FBKV{}.
Although Aeolous' credit-based rate control for non-first RTT traffic reduces buffer occupancy, 
it introduces a new challenge: when a sender has credits but has no data to be sent out in a timely manner, the allocated bandwidth will be wasted instead of giving to other senders~\cite{Homa-Sigcomm18}. \FBKV\ has lighter load with smaller messages, causing this problem to happen more often.

%


Less intuitively, \sysname\ also outperforms UDP. This is because 1) UDP unnecessarily creates congestion in the network,
and 2) when loss happens, \sysname\ can retransmit unacknowledged packets if necessary, while UDP has to wait for new packets to be produced by the application. 
Moreover, UDP does not provide accuracy guarantees.
We observe that UDP's loss rate is 35\%, while \sysname's loss rate is significantly lower (8.8\%).


To better understand \sysname's performance improvements, 
we examine the queuing effect of \sysname\ and the other four schemes in Figures~\ref{fig-queue-fbkv} and \ref{fig-queue-fbhd}. 
As expected, \sysname\ has the smallest queue length (less than 1 MTU).
pFabric causes significant queuing, and the queue size grows as MSR gets higher.
This is because higher MSR implies more data will be sent and pFabric always starts each flow by sending at line rate, which results in significant queuing, congestion and ultimately timeouts.
While not as severe as pFabric, DCTCP also has high queueing 
because its queue can build up when multiple flows compete for a switch output port, \ie, incast -- a well studied drawback of DCTCP~\cite{ndp-sigcomm17, Homa-Sigcomm18}.
UDP's queuing is high, especially with \FBHD\ (for which UDP is similar to pFabric), because UDP also sends at the line rate and \FBHD\ has a more intensive data arrival rate than \FBKV.
Aeolous has a much smaller queue length than the other non-ATP schemes, because after the first RTT, senders only send messages based on credits received from the receiver, which allows Aeolous to control queuing size. 

In addition to the all-to-all traffic pattern, we evaluate an all-to-one pattern, i.e., incast, where all hosts in the network send to the same host.
Figure~\ref{fig-all-to-one} shows the JCT of \sysname, DCTCP-SS, and DCTCP-ES.
We do not include pFabric or Aeolous results in this figure because their performance is much worse than DCTCP's (in fact, Aeolous fails to complete the test).
In particular, the incast pattern introduces a bottleneck and significant queuing at a specific Top-of-Rack switch. With pFabric, we observed that this incast resulted in significant queue build up and ultimately triggered timeouts which impacted performance. While Aeolous was able to avoid this level of build-up with its credit based mechanisms, we observed that Aeolous has slow performance because of a significant amount of credit waste. Digging deeper, we observed that a combination of the short flows and inter-arrival times of FBKV resulted in scenarios were some flows received credits but were unable to use the credits.
\sysname\ limited buffer allocations prevents packet build up but incurs loss which approximate applications can tolerate and \sysname\ exploits by avoiding re-transmissions and delivering packets out of order. 


{
\begin{figure*}[th]
\begin{minipage}{0.245\textwidth}
\begin{center}
\centerline{\includegraphics[width=1.0\columnwidth]{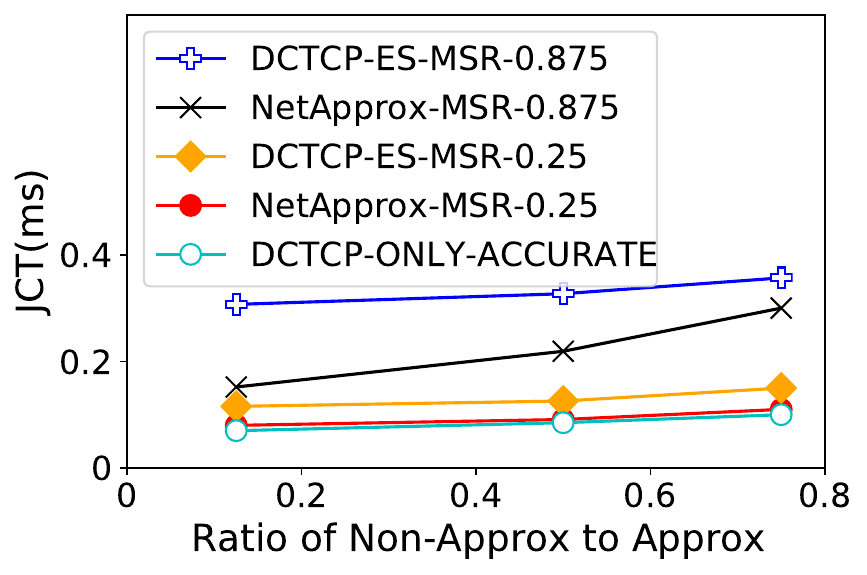}}
\vspace{-0.1in}
\mycaption{fig-non-approx}{Effect on Non-Approximate Traffic.}
{
}
\end{center}
\end{minipage}
\begin{minipage}{0.245\textwidth}
\begin{center}
\centerline{\includegraphics[width=0.95\columnwidth]{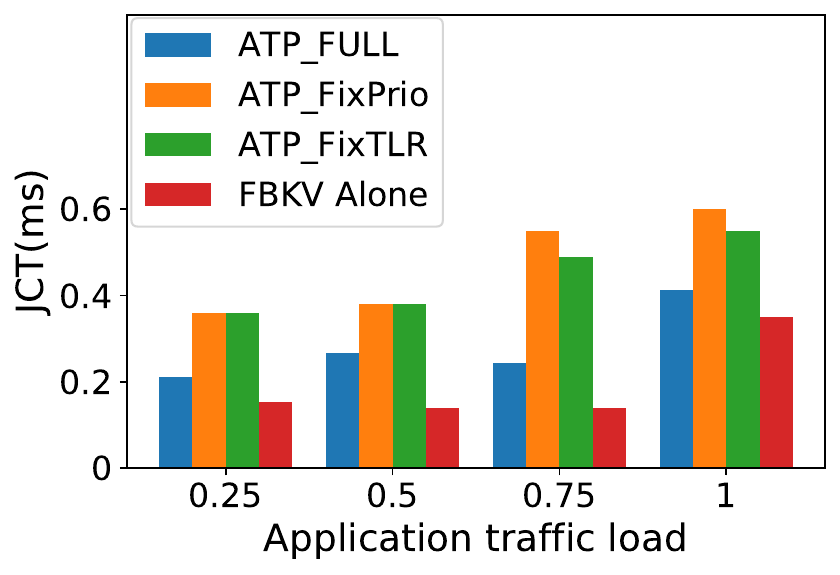}}
\vspace{-0.1in}
\mycaption{fig-fairness}{Job Fairness.}
{
FBKV JCT when two FBHD jobs are competing.
}
\end{center}
\end{minipage}
\begin{minipage}{0.245\textwidth}
\begin{center}
\centerline{\includegraphics[width=1.0\columnwidth]{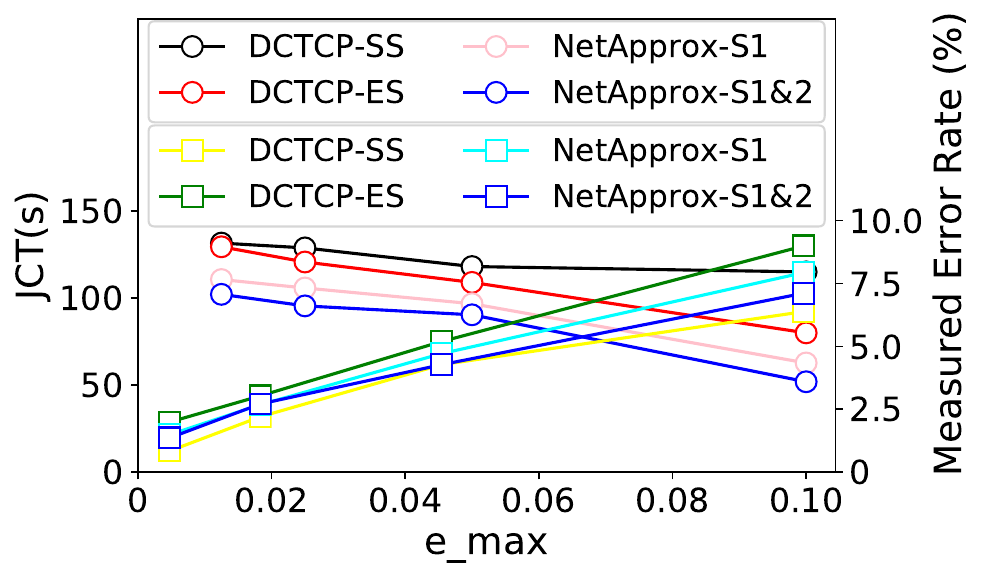}}
\vspace{-0.1in}
\mycaption{fig-spark-batch}{Spark batch job JCT and error rate.}
{
lines with circle represent \jct, lines with square represent 
measured error rate.
}
\end{center}
\end{minipage}
\begin{minipage}{0.245\textwidth}
\begin{center}
\centerline{\includegraphics[width=0.9\columnwidth]{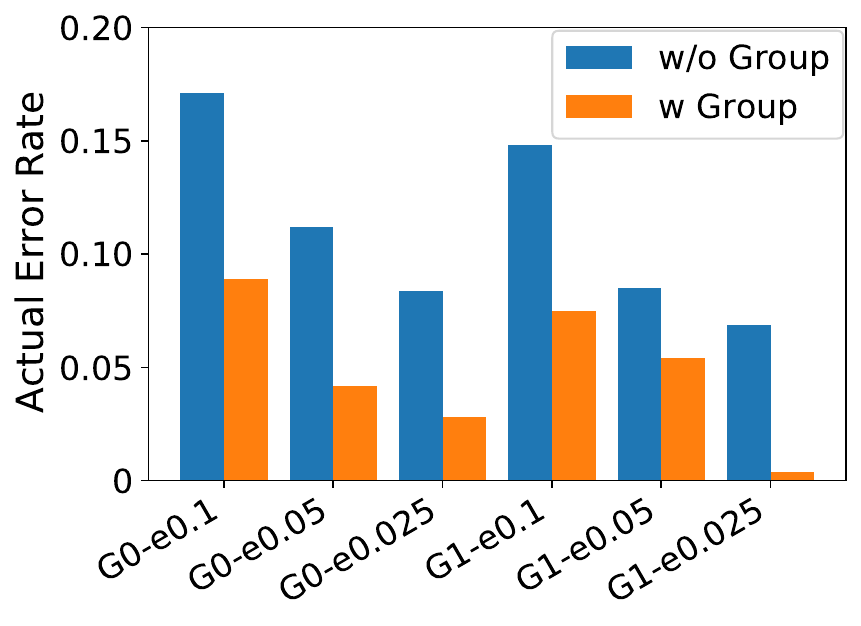}}
\vspace{-0.1in}
\mycaption{fig-spark-groupby}{Group effect on Spark job.}
{\eg, G0-e0.1 means Group-0 with $e_{max}$ 0.1. 
}
\end{center}
\end{minipage}
\vspace{-0.2in}
\end{figure*}
}

\subsubsection{Dynamic Adjustment of \tlr\ and Priority}
A key technical contribution of \sysname\ is its ability to dynamically adjust the target loss rate (and as a result, the sending rate) and priority based on network states and a job's changing slack tolerance.
To evaluate the effectiveness of \sysname's dynamic mechanisms, we create synthetic, controlled background traffic, while running the foreground job of \FBKV\ with \msr\ 0.5.
Figure~\ref{fig-dynamic-timeline} shows the timeline of how measured loss rate, \tlr, sending rate, and priority assignment change for the foreground job.
In the beginning, the foreground job is assigned to switch queue4 (because its initial TLR is 0.5).
We create the background traffic to initially also go to queue4.
Because of the competing background traffic, the foreground job's actual loss rate gets higher than its \tlr\ even when \atp\ quickly slows down the sending rate.
\atp\ then reduces its \tlr, its sending rate, and shifts the job to queue3 (one priority higher than queue4).
The job's loss rate then drops (since there is no other competing traffic in queue3) and stays close to its \tlr;
as a result, it sends at a more aggressive rate.
Then at time 0.59, we shift the background traffic to queue 3 which starts to compete with the foreground job again.
With the loss rate again higher than \tlr, \atp\ adjusts the \tlr\ and moves the job to queue2, which then results in the drop of loss rate.
Towards the end of the test, we shift the background traffic to queue2, and \atp\ adjusts the \tlr\ and moves the job to queue1.

To understand how effective our dynamic mechanisms are, we perform a set of experiments similar to the one above, 
\ie, by having background traffic that starts at the same queue as the foreground job's and change to one priority higher in the half way.
For the foreground job, we use \FBKV\ with different \msr{}s (thus, they fall into different priority queues initially).
Figure~\ref{fig-dynamic-overall} plots the JCT of four schemes: DCTCP-ES,
\atpfixtlr\ which statically sets \tlr\ to $1-MSR$,
\atpfixqueue\ which dynamically sets \tlr\ but never moves the job from its initially assigned queue,
and \atpfull\ which includes both the two dynamic settings.
As expected, \atpfull\ performs the best, and both dynamic mechanisms improve JCT.
Dynamic setting priority (\ie, the difference between \atpfixqueue\ and \atpfull) has huge improvements on JCT, especially for small \msr.
This is because \atpfull\ can use a much higher sending rate (esp. for smaller \msr) after it shifts the foreground traffic to higher-priority queues and avoids being in the same queue as the background traffic.
Dynamic setting of \tlr\ (\ie, the difference between \atpfixtlr\ and \atpfixqueue) is more effective with larger \msr, because a not-so-well-set \tlr\ as in \atpfixtlr\ could demand more retransmission to achieve a higher \msr\ than when the \msr\ is smaller and more slack can be tolerated.
Finally, even without dynamic mechanisms, \atp\ still outperforms DCTCP-ES because of its approximate-aware rate control.

\subsubsection{Effect of \atp\ Techniques}
To further understand where \atp's performance gain comes from, 
we dissect the effect of \atp's various components besides dynamic \tlr\ and priority-queue setting.
We compare (1) \atpbase, a base protocol that uses raw UDP and after sending out all packets, retransmit lost packets until $MSR \times\ N$ messages are delivered,
(2) \atprc, which adds approximate-aware rate control to \atpbase\ with \tlr\ statically set to $1-MSR$,
(3) \atpsched, which adds the MRDF scheduling policy on top of \atprc,
and (4) \atpfull, the final \atp\ protocol with dynamic \tlr\ and priority settings.
Figure~\ref{fig-atp-techniques} shows the \jct\ of these schemes when running the \FBKV\ workload with different \msr.
As expected, rate control, \ie, \atprc, largely improves the basic protocol (by up to 42\%);
the inclusion of our scheduling algorithm, \ie, \atpsched, improves performance by up to 39\% over \atprc;
and our dynamic mechanisms, \ie, \atpfull, further improves performance by up to 47\%.
When \msr\ is small, the effect of different techniques is not obvious, as there is more slack tolerance with small \msr\ and a simple technique could just work when there is no competing traffic in the network.

\subsubsection{Impact on Non-Approximate Traffic}
\label{sec:accurate}

One of the key benefits of \sysname\ is in its improvement of non-approximate traffic performance when co-running with approximate traffic.
To demonstrate this benefit, we change the ratio of \nonapproximate\ and approximate traffic
and run the approximate traffic (\FBKV{}) using \atp\ and DCTCP-ES, with different \msr{}s.
As shown in Figure~\ref{fig-non-approx}, \atp\ largely improves \nonapproximate{}'s \jct\ for all the \msr{}s and ratios, especially when there are more approximate traffic in the network and when \msr\ is larger.

\subsubsection{Job-Level Fairness}

A key benefit of \sysname's slack-based switch priority queue design is the avoidance of starvation, \ie, workloads with more slack (and thus more aggressive sending rate) should not starve workloads with little or no slack.
To evaluate this design, we perform an experiment that runs three approximate jobs together: a \FBKV\ job with $MSR=0.875$, a \FBHD\ job with $MSR=0.75$, and a \FBHD\ job with $MSR=0.5$.
The \FBKV\ job starts after the two \FBHD\ jobs, and the \FBHD\ workload is more intensive than \FBKV{}.

{
\begin{figure}[th]
\begin{center}
\begin{minipage}{0.22\textwidth}
\begin{minipage}{0.96\columnwidth}
\begin{center}
\centerline{\includegraphics[width=1.0\columnwidth]{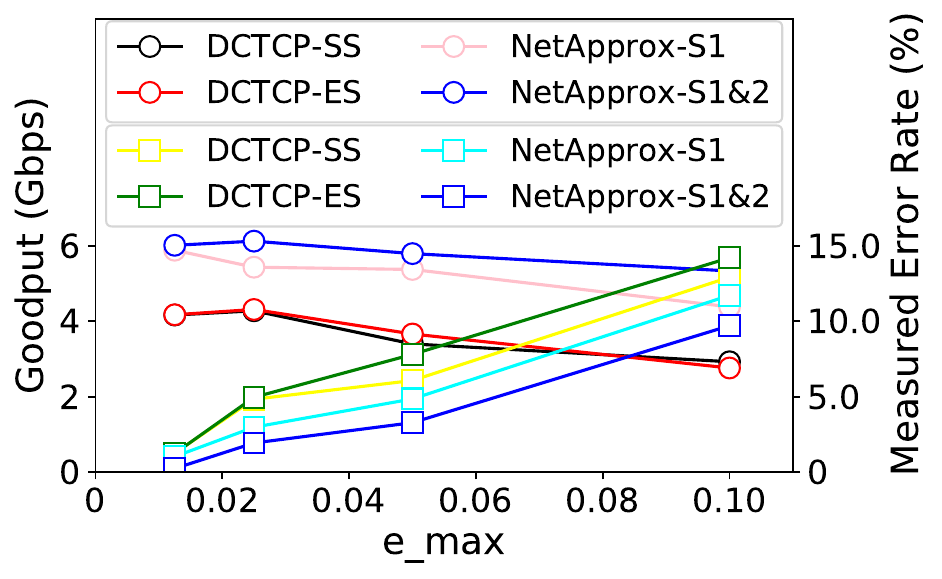}}
\end{center}
\end{minipage}
\vspace{-0.3in}
\mycaption{fig-flink-streaming}{Flink streaming goodput and error rate.} 
{
lines with circle represent goodput, lines with square represent 
measured error rate.
}
\end{minipage}
\begin{minipage}{0.23\textwidth}
\begin{minipage}{0.96\columnwidth}
\begin{center}
\centerline{\includegraphics[width=1.0\columnwidth]{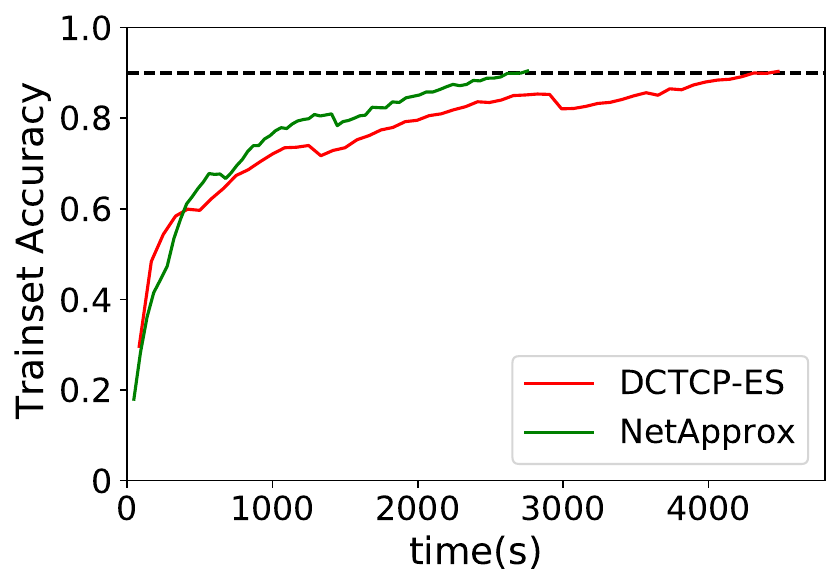}}
\end{center}
\end{minipage}
\vspace{-0.3in}
\mycaption{fig-pytorch-trainset}{VGG19 training set convergence.}
{ 
}
\end{minipage}
\end{center}
\end{figure}
}

Figure~\ref{fig-fairness} plots the JCT of the \FBKV\ workload under four schemes. 
\atpfull\ effectively prevents the light \FBKV\ workload from being affected
by data intensive traffic \FBHD\ and achieves the closest performance as when running \FBKV\ alone, since it moves the \FBKV\ traffic to a higher priority queues.
\atpfixqueue\ and \atpfixtlr\ both result in worse \FBKV\ performance, because they do not move the traffic to another priority queue. \atpfixqueue\ performs worse than \atpfixtlr\ for lighter \FBKV\ traffic load, \eg, 0.25, it is because \atpfixqueue\ keeps decreasing its rate when heavy traffic occupies the same queue to minimize the 
retransmissions.



\subsection{Real Implementation Results}
\label{sec:real-results}

\boldparagraph{Environments.} 
We evaluate \sysname\ on a lab cluster with five servers 
and one 100\gbps, 32-port N8560-32C Ethernet switch. 
Each server has two 12-core CPU and 64\GB\ memory.
Three servers are equipped with one Nvidia A6000 GPU respectively.
Our switch is configured to use 32 \MB\ shared buffer for all the 32 ports. 
%
To model a real switch's load, we generate background traffic according to production data-center network traffic distributions~\cite{Microburst-imc17}.

\subsubsection{Spark Results}

We use four servers as spark workers and three servers as kafka broker which are the data sources.
We use the NYT taxi dataset~\cite{taxi-trace} for our Spark and Kafka/Flink experiments. It consists of itinerary information of all rides across yellow and green taxi’s in New York City from year 2017 and 2018. The total volume of raw data is about 21\GB.
We evaluated two workloads on Spark: \avg\ of hourly ride distance of the NYT trace, \rank\ to find the top 15 taxi pickup locations within every hour of the NYT trace.
Because space constraints, we only present the result of \avg;
the \rank\ results are qualitatively similar. 

\boldparagraph{Sampling without group.}
We first use simple sampling for these jobs, 
\ie, \sysname\ treats all messages as coming from a single group with the same importance.
Users specify an $e_{max}$, which \sysname\ iteratively adjusts its \msr\ to meet. Once it is met, the job is considered finished.
Figure~\ref{fig-spark-batch}
shows the \jct\ and error rate of the \avg\ Spark job 
with different $e_{max}$ using DCTCP-ES, DCTCP-SS, \sysname\ applied only at the input stage (denoted by \sysname-S1), 
and \sysname\ applied at both the input and mapper-to-reducer stages (denoted by \sysname-S1\&2). 

Performing approximation with \sysname\ largely improves Spark jobs' overall performance while only incurring less than 1\%\ accuracy reduction.
\sysname\ outperforms DCTCP-SS and DCTCP-ES by 21.1\% to 54.8\%.
This is because when switch ports are congested (with background traffic), \dctcp\ incurs excessive timeouts and retransmissions, while \sysname\ adapts to congestion well with our slack-tolerance transport. 
As expected, using \sysname\ for both stages improves \jct\ over when applying \sysname\ only for input stage.
\sysname\ also achieves better accuracy than DCTCP-ES, because DCTCP-ES only sends the exact amount to reach user-specified $e_{max}$ and stops. In contrast, \sysname\ can send more data if doing so would not affect JCT (when the network has extra room). As a result, \sysname\ can achieve higher accuracy than $e_{max}$.
\sysname's accuracy is slightly worse than DCTCP-SS because DCTCP-SS performs exact uniform sampling, which results in the best sampling quality for this problem. Our sampling is not as uniform to trade for better performance, while still controlling errors within $e_{max}$.

\boldparagraph{Sampling with groups.}
The above results treat all data points the same when sampling.
As discussed in \S\ref{sec:group-sampling}, users can have many different sampling criteria for which we offer the {\em group} semantics.
To demonstrate this usage, we perform a Spark data analytics that computes the \avg\ of data with two different keys (\ie, using \texttt{ReducebyKey}), as shown in Figure~\ref{fig-spark-groupby}.
When using groups, \sysname\ achieves the same sampling quality within each key, thereby improving the overall accuracy. 
Without groups, \sysname\ samples across all the data from all keys and can end up not meeting the $e_{max}$ requirement within each key.


\subsubsection{Kafka and Flink Results}
We run three Kafka producers, three Kafka brokers, and four Flink workers. Each producer generates some input data and sends it to a broker in a streaming way. The brokers forwards the received data to the Flink workers.
We run a window-based workload on this platform which calculates the \avg\ of ride distance data received in every one-minute window using the NYT taxi dataset.
Since streaming jobs like this do not have a fixed ``job'' or JCT, we report the average goodput achieved during the streaming.

Figure~\ref{fig-flink-streaming} plots the average goodput and measured accuracy with \sysname\ applied only to the Kafka-Flink-data-feeding stage, \sysname\ applied to both this stage and the data-processing stage within Flink, DCTCP-ES, and DCTCP-SS.
Similar to the Spark results, \sysname\ outperforms both DCTCP-SS and DCTCP-ES (higher goodput), and applying \sysname\ to both stages further improves goodput.
For this workload, \sysname's accuracy is better than both DCTCP-SS and DCTCP-ES.

\subsubsection{PyTorch Results}

We use the three GPU-equipped servers in our cluster to run distributed DNN training on PyTorch. 
We train the VGG19 model~\cite{vgg19} on the CIFAR-10 dataset~\cite{Cifar10} with minibatch size 80, learning rate 0.00002, cosine annealing, and Adam optimizer.
Figure~\ref{fig-pytorch-trainset} shows the convergence timeline with X-axis as real wall-clock time and Y axis as the training-set accuracy when using \sysname\ and DCTCP-ES.
For \sysname, we use a group policy where gradients less than 0.001 have \msr\ 0, \ie, can all be dropped,
gradients between 0.001 and 0.0025 have \msr\ to 0.125,
and all gradients above 0.0025 have \msr\ to 0.925.
We obtained these hyper-parameters from few rounds of tuning. We also tested another simpler group policy, where all gradients below 0.003 have \msr\ 0 and all other ones have \msr\ 1; the results are only slightly worse than the more complex group policy, demonstrating the robustness of approximation.

As seen, \sysname\ reaches 90\% training-set accuracy at 46 minutes, while DCTCP-ES takes 75 minutes to reach the same accuracy.
The test-accuracy convergence (not shown for space reason) exhibit similar trends: \sysname\ reaches the same test-set accuracy 45\% faster than \dctcp-ES. 

\section{Conclusion}
\label{sec:conclude}

This paper presents  \sysname, the first network system that is designed for approximate computing.
\sysname\ leverages the inherent slack in approximate applications by embracing loss in the network 
and by assigning tiny switch resources to approximate traffic.
Our large-scale simulation and real implementation evaluation demonstrate that \sysname\ simultaneously improves both approximate and \nonapproximate\ applications' performance while guaranteeing user-specified accuracy requirements.

\newpage

\bibliographystyle{ACM-Reference-Format}
\bibliography{all-defs,all,personal,all-confs,local,paper}

\end{document}